\documentclass[prd,aps,twocolumn,a4paper,floatfix,amsmath,amssymb,nofootinbib]{revtex4-1}
\usepackage{graphicx}
\newcommand{\bp}{{\bar p}}
\newcommand{\bq}{{\bar q}}
\newcommand{\tprop}{t}

\newif\ifprintfigures
\printfigurestrue 

\begin{document}

\title{Critical gravitational collapse with angular momentum II: soft
  equations of state}

\author{Carsten Gundlach}

\affiliation{Mathematical Sciences,
University of Southampton, Southampton SO17 1BJ, United Kingdom}

\author{Thomas W. Baumgarte}

\affiliation{Department of Physics and Astronomy, Bowdoin College,
  Brunswick, ME 04011, USA}

\date{24 January 2018}


\begin{abstract}
We study critical phenomena in the collapse of rotating
ultrarelativistic perfect fluids, in which the pressure $P$ is related
to the total energy density $\rho$ by $P=\kappa\rho$, with $\kappa$ a
constant.  We generalize earlier results for radiation fluids with
$\kappa=1/3$ to other values of $\kappa$, focussing on $\kappa <
1/9$. For $1/9<\kappa \lesssim 0.49$, the critical solution has only
one unstable, growing mode, which is spherically symmetric. For
supercritical data it controls the black hole mass, while for
subcritical data it controls the maximum density.  For $\kappa<1/9$,
an additional axial $l=1$ mode becomes unstable. This controls either
the black hole angular momentum, or the maximum angular velocity.  In
theory, the additional unstable $l=1$ mode changes the nature of the
black hole threshold completely: at sufficiently large initial
rotation rates $\Omega$ and sufficient fine-tuning of the initial data
to the black hole threshold we expect to observe nontrivial universal
scaling functions (familiar from critical phase transitions in
thermodynamics) governing the black hole mass and angular momentum,
and, with further fine-tuning, eventually a {finite} black hole mass
almost everywhere on the threshold.  In practice, however, the second
unstable mode grows so slowly that we do not observe {this breakdown
  of scaling at} the level of fine-tuning we can achieve, nor
systematic deviations from the leading-order power-law scalings of the
black hole mass. We do see systematic effects in the black-hole
angular momentum, but it is not clear yet if these are due to the
predicted non-trivial scaling functions, or to nonlinear effects at
sufficiently large initial angular momentum (which we do not account
for in our theoretical model).

\end{abstract}


\maketitle

\tableofcontents


\section{Introduction}

Since the pioneering numerical work of Choptuik \cite{Choptuik1993} on
the massless scalar field, Evans and Coleman \cite{EvansColeman} on
the perfect fluid, and Abrahams and Evans \cite{AbrahamsEvans} on
vacuum collapse, it has been know that interesting things happen at
the {\em black hole threshold}: the codimension-one hypersurface in
the space of regular initial data that separates {supercritical} data
which eventually form a black hole from {subcritical} data which do
not.

What happens there can be summarized as universality, a self-similar
contraction phase during the time evolution, and power-law
scaling of quantities such as the black hole mass with distance to the
black hole threshold. An example of this is the famous formula for the
black hole mass $M$ {for supercritical data,}
\begin{equation} \label{choptuikscaling}
M(p) \sim (p - p_*)^\gamma,
\end{equation}
where $p$ is the parameter of a generic two-parameter family of
initial data, $p = p_*$ is the critical parameter that identifies the
black hole threshold (we have assumed for definiteness
  that a black hole forms for $p>p_*$), and $\gamma$ is the critical
exponent.  We refer the reader to \cite{LRR} for a review and further
references to the literature.

In this paper we extend our numerical study of the critical collapse
of rotating perfect fluids with the linear, ultrarelativistic equation
of state $P=\kappa \rho$ from the radiation gas case $\kappa=1/3$
(see, e.g., \cite{EvansColeman,BaumgarteGundlach,gb16}) to other
values of $\kappa$, in particular to values $0<\kappa<1/9$ (see also
  \cite{NeilsenChoptuik} for a study of such fluids in
    spherical symmetry).  We expect this to be interesting as then
the spherically symmetric critical solution
\cite{KoikeHaraAdachi,Maison} has not one but two growing modes
\cite{critfluidpert}: the familiar spherical one, plus an axial $l=1$
mode that can be thought of as a spin-up under self-similar
contraction. For $\kappa \gtrsim 0.49$ an $l = 2$ mode is expected to
become unstable \cite{critfluidpert}, but in this paper we focus on
the $l=1$ modes for rotating fluids.

To understand the relevance of a second unstable mode, first go back
to initial data restricted to spherical symmetry. The critical
solution then has only one unstable mode and hence, assuming
linearization stability, it must have a finite-sized attracting
manifold of codimension 1. In our matter model, all initial data
either produce a black hole or disperse. Define the black hole
threshold in phase space as the hypersurface (codimension-1 manifold)
which separates the data that collapse from {those} that disperse. As
the critical solution neither collapses nor disperses (it is
self-similar and produces a naked singularity), its initial data must
lie on the black hole threshold, and hence the black hole threshold
and the attracting manifold of the critical solution coincide at least
near the critical solution. In fact, numerical results in spherical
symmetry indicate that the attracting manifold of the critical
solution is the entire black hole threshold.

For generic spherically symmetric initial data sufficiently near the
black hole threshold, there are then three phases of the time
evolution. In Phase~1, the solution moves a {potentially} long
distance from its starting point to near the critical solution. In
Phase~2, the solution is approximated by the critical solution
  $Z_*$ plus linear perturbations. It leaves this phase when the
{amplitude of the} single growing perturbation mode $Z_0$ has grown
sufficiently large.  We normalize $Z_0$ so {that} this occurs at
amplitude $\pm 1$.  The moment when this happens sets an overall
length scale: as $Z_*$ is a self-similar contraction, the
later $Z_0$ becomes nonlinear, the smaller the scale.  In Phase~3 the
evolution is again non-linear, but as the intermediate Cauchy data at
the start of Phase~3 are universal up to the overall scale, namely
$Z_*\pm Z_0$, this third phase is also universal up to this overall
scale and the sign of $Z_0$: $Z_*+Z_0$ evolves into a black hole,
while $Z_*-Z_0$ disperses.

If we now consider generic initial data that are near the black hole
threshold and slightly non-spherical, and if all non-spherical modes
are stable {(i.e.~for $\kappa > 1/9$ in our matter model)}, then
essentially the same picture holds: Deviations from spherical symmetry
in Phase~1 are small and can be treated as linear perturbations simply
because they are small in the initial data. They decay further in
Phase~2. At the beginning of Phase~3 we {now} have data $Z_*\pm Z_0 +
\delta\, Z_1$, where $Z_1$ is the most slowly damped axial $l=1$
perturbation mode, and by assumption its amplitude $\delta$ is
small. In Phase~3, angular momentum can then be treated as a linear
perturbation of the nonlinear evolution of the spherical data $Z_*\pm
Z_0$, right up to the formulation of a slowly rotating black hole
(considered as a linear perturbation of Schwarzschild). We normalize
$Z_1$ so that $J/M^2$ of the black hole equals $\delta$. Because $Z_1$
is a decaying mode, $\delta\to 0$ as the initial data are fine-tuned
to the black hole threshold, and so the perturbative treatment of
angular momentum is consistent in this limit. Our previous work on
$\kappa=1/3$ bore out this theory \cite{BaumgarteGundlach,gb16}, but
we also found that it made quantitatively correct predictions up to
fairly large values of $J/M^2$ -- an example of the ``unreasonable
effectiveness'' of perturbation theory.

Consider now the case where both $Z_0$ and $Z_1$ are growing
modes (and there are no others {-- i.e.~for $\kappa < 1/9$ in our
  matter model)}. Consider again initial data whose deviation from
spherical symmetry is small enough to remain linear in
Phase~1. However, the amplitude of $Z_1$ now grows during
Phase~2. Therefore, its amplitude at the end of Phase~2 can become
large as we fine-tune the initial data to the black hole threshold. In
\cite{scalingfunctions} one of us concluded that the final outcome of
Phase~3 would now depend not only the sign in front of $Z_0$, but also
on the value of $\delta$. {In particular, we expect that a black
  hole forms for $|\delta|<\delta_*$ (although the threshold value
  $\delta_*$ may be infinite) and that}
$J/M^2=F_{J/M^2}(\delta)$ for some non-trivial ``universal scaling
function'' $F_{J/M^2}$ \cite{scalingfunctions}.  To linear order we
still expect $F_{J/M^2}(\delta) \simeq \delta$, but when $Z_1$ is a growing
mode, $\delta$ can become large, in which case higher-order terms in
$F_{J/M^2}(\delta)$ affect $J/M^2$.

Moreover, if we fine-tune any one parameter in the initial data to the
black hole threshold, we will not fine-tune the initial amplitudes of
both growing modes to zero, reaching some minimum {initial
  amplitude} somewhere near the black-hole threshold. Hence we
formally expect to see {a breakdown of scaling sufficiently
  close to the black-hole threshold}. However, we cannot at the moment
exclude the possibility that the black hole mass and spin go to zero
at the black hole threshold anyway, since this might still happen if
the universal scaling functions $F_M(\delta)$ and $F_J(\delta)$ for
the mass and spin vanish {at the black hole threshold given by
  $\delta=\delta_*$}.

With two growing modes, in principle we need to fine-tune two
parameters in the initial data in order to make the initial amplitudes
of $Z_0$ and $Z_1$ small. However, any initial data which have a
reflection symmetry cannot have angular momentum, and so $Z_1$, which
is associated with angular momentum, cannot arise. Hence if we choose
one of our two parameters (call it $q$) such that the initial data
have a reflection symmetry for $q=0$ but not for $q\ne 0$, then we
know a priori that setting $q=0$ in the initial data will set the
amplitude of $Z_1$ to zero, and we only need to fine-tune the other
parameter (call it $p$) to also set the amplitude of $Z_0$ to zero
\cite{scalingfunctions}.\footnote{{Another matter model that admits a
    2-mode-unstable critical solution, even in spherical symmetry, is
    a harmonic map coupled to gravity \cite{Liebling1998}.  In this
    model,} the amplitude of one of the two growing modes can be set
  to zero a priori by maximizing charge density over mass density in
  the initial data.}

\section{Theory}
\label{sec:theory}

\subsection{The critical solution}

For the perfect fluids considered here, the critical solution is
continuously self-similar.  The only physical length scale of this
solution can then be expressed as $\tprop_* - \tprop$, where $\tprop$
is the proper time of an observer at the center, and $\tprop_*$ refers
to the accumulation event, i.e.~the instant of proper time at which
the solution has contracted to zero radius.  Any quantity of dimension
$L^n$, where $L$ represents a length scale, must therefore scale with
$(\tprop_* - \tprop)^n$.  Any non-dimensional quantity becomes
time-independent when expressed in terms of spatial coordinates $x$
that are dragged along with the self-similar contraction.
Symbolically, we may therefore write non-dimensional quantities
associated with the critical solution as $Z_*(x)$.  Since the only
length scale $\tprop_* - t$ depends on time, the critical solution
evidently does not display any global length scale and is therefore
completely scale-invariant, or continuously self-similar.  The
length scale that determines, for example, the mass of black holes
formed in supercritical evolutions is determined by the time
  when the solution starts to deviate significantly from the critical
  solution, and hence by the growing perturbations of the critical
solution.

\subsection{Spherically symmetric {one}-parameter families of
    initial data}
\label{sec:theory:ss}

We first focus on spherical symmetry and assume that the
{(non-rotating)} initial data are parameterized by a single parameter
$p$.  In spherical symmetry, the critical solution has exactly one
growing mode, which we denote by $Z_0(x)$.  Singling out this growing
mode, we may approximate the solution in Phase~2 (near
self-similarity) as
\begin{equation} \label{approx_1} 
Z(\tprop,x) = Z_*(x) + \zeta_0(t) Z_0(x) + \ldots
\end{equation}
Linear perturbations that grow on a constant time scale increase
exponentially in the time $\tprop$.  Here, however, the time scale of
growth is the scale of the self-similar solution, $\tprop_* - \tprop$,
so that the amplitude $\zeta_0$ of the perturbation satisfies an
equation of the form
\begin{equation}
\frac{d \zeta_0}{dt} \propto \frac{\zeta_0}{\tprop_* - \tprop}.
\end{equation}
Accordingly, the mode grows exponentially in the dimensionless time
coordinate
\begin{equation} \label{tau_def}
\tau := - \ln {\tprop_* - \tprop\over c_t L_0},
\end{equation}
where $L_0$ is an arbitrary fixed length scale, and $c_t$ a
dimensionless universal constant. (In the following, we measure all
dimensional quantities in units of $L_0$.) We may therefore write
\begin{equation}
\zeta_0 = P(p) e^{\lambda_0 \tau},
\end{equation}
where the factor $P(p)$ depends on the initial data, and where
$\lambda_0$ is the Lyapunov exponent of the one unstable spherical
perturbation.  Since $P(p)$ must vanish for the critical solution
parametrized by $p=p_{*0}$, where we have inserted a subscript $0$
referring to zero rotation, we may expand to leading order as
\begin{equation} \label{P_expansion_1}
P(p) = C_0 (p - p_{*0}) + \ldots
\end{equation}
The constant $C_0$ depends on the particular one-parameter family of
initial data and its parameterization.  For definiteness, we assume
that a black hole forms for $P>0$.

The scale of dimensional quantities formed in near-critical collapse
is determined by the time at which perturbations become nonlinear,
i.e.~when the evolution transitions from Phase 2 to Phase 3.  We
normalize the linear perturbation mode $Z_0(x)$ so that the growing
mode becomes nonlinear when $\zeta_0 = \pm 1$. This occurs at a
time\footnote{We note that $\tau_\sharp$ was called $\tau_*$ in our
  previous papers.}
\begin{equation}
\tau_\sharp := -{\ln |P(p)| \over \lambda_0}.
\end{equation}
The corresponding length scale is 
\begin{equation} \label{length_scale}
e^{-\tau_\sharp} = {\tprop_* - \tprop_\sharp \over c_t} =  
|P(p)|^{1\over\lambda_0}.
\end{equation}
In supercritical evolutions, the mass of the black hole that then
forms must be proportional to this length scale, or more precisely
\begin{equation} \label{mass_scaling_2}
M \simeq c_M \left(C_0(p - p_{*0})\right)^{1\over\lambda_0},
\end{equation}
where we have inserted the expansion (\ref{P_expansion_1}), and where
$c_M$ is a dimensionless universal constant. Hence the critical
exponent $\gamma_M$ for the black hole mass in
  (\ref{choptuikscaling}) is the inverse of the growth rate
$\lambda_0$,
\begin{equation}
\gamma_M = {1\over\lambda_0}.
\end{equation}
By shifting the origin of $\tau$, we set $c_M=1$ in the
following. (This leaves $c_t$ as a dimensionless universal constant
that we cannot set to one.)

\begin{figure}
\includegraphics[width=3.5in]{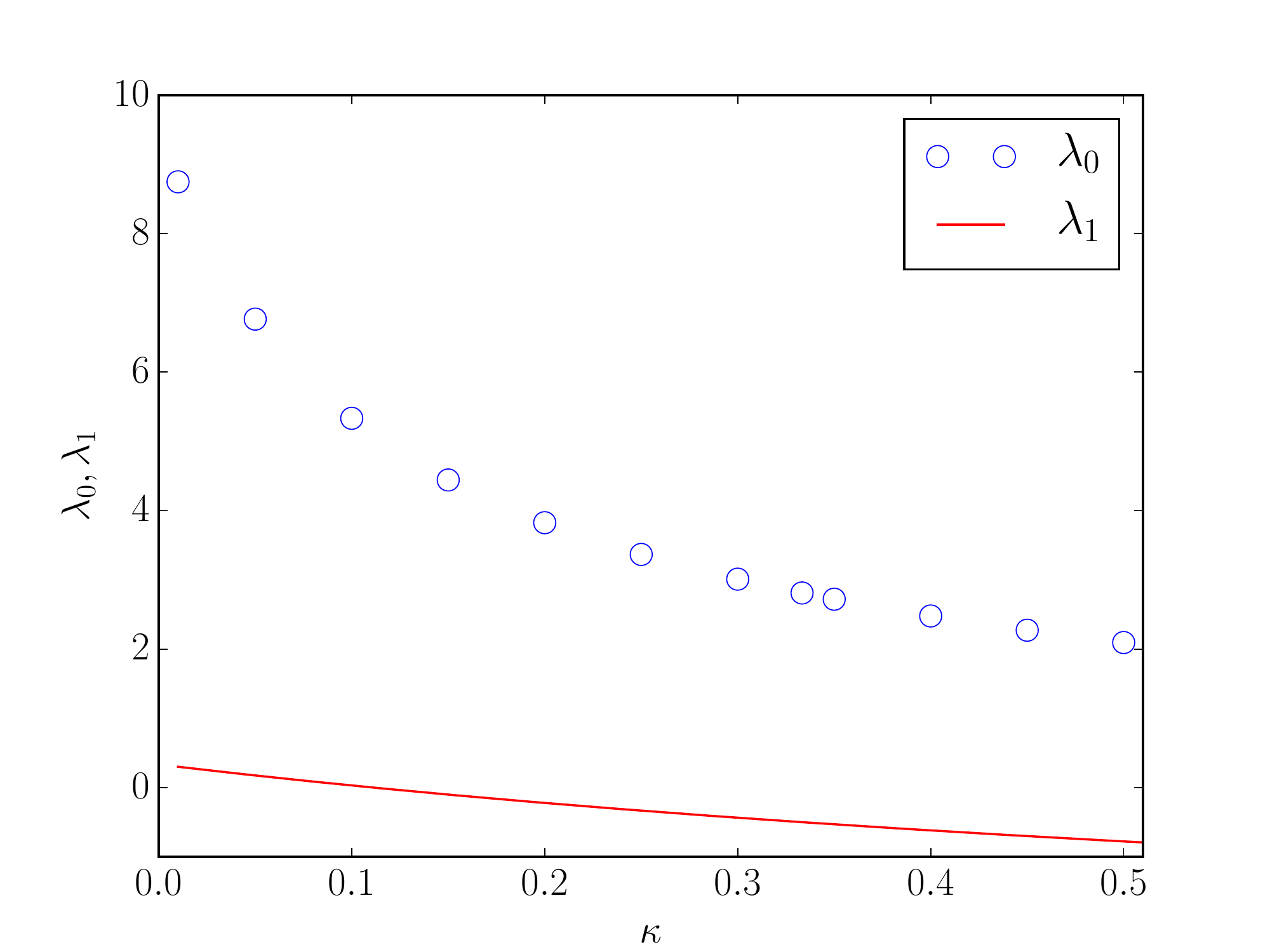}
\caption{Perturbative values of the Lyapunov exponents $\lambda_0$ and
  $\lambda_1$ for perfect fluids with the equation of state
  (\ref{eos}).  The values for $\lambda_0$ are taken from
  \cite{Maison}, while the values for $\lambda_1$ are given by
  (\ref{lambda1theory}).  The exponent $\lambda_1$ changes sign at
  $\kappa = 1/9$, reflecting the fact that the $Z_1$ mode becomes
  unstable for $\kappa < 1/9$.  The values of $\lambda_0$ are much
  greater than those of $\lambda_1$ in this regime, which explains why
  the effects of $Z_1$ becoming unstable are difficult to observe in
  numerical simulations.}
\label{Fig:lambdas}
\end{figure}

In this paper we focus on perfect fluid matter with the linear,
ultrarelativistic equation of state
\begin{equation} \label{eos}
P = \kappa \rho,
\end{equation}
where $P$ is the pressure, $\rho$ the total energy density, and
$\kappa$ a constant.  A radiation fluid, in particular, is described
by $\kappa = 1/3$.  For such fluids, and for a given value of
$\kappa$, the Lyapunov exponent $\lambda_0$ can be computed by
perturbing the critical solution (see, e.g., \cite{Maison}, as well as
Fig.~\ref{Fig:lambdas}.)

It was first observed in \cite{GarfinkleDuncan} that critical scaling
occurs in sub- as well as supercritical evolutions. In our matter
model, by dimensional analysis, the maximum density attained in
subcritical evolutions must scale with the inverse square of the
length scale (\ref{length_scale}), so that for subcritical evolutions
we may write.
\begin{equation}
\rho_{\rm max} \simeq c_\rho\left(C_0(p_{*0}-p)\right)^{\gamma_\rho}
\end{equation}
with 
\begin{equation}
\gamma_\rho =  -{2\over \lambda_0},
\end{equation}
and $c_\rho$ another dimensionless universal constant.

\subsection{Two-parameter families of rotating initial data}

We now consider rotating initial data, and assume that these data are
analytic and parametrized by two parameters $p$ and ${\bf q}$.  We
further assume that, if these data evolve to form a black hole, the
black hole mass $M$ and angular momentum ${\bf J}$ obey the symmetries
\begin{subequations} \label{symm}
\begin{eqnarray}
\label{psymm}
M(p,- {\bf q})&=&M(p,{\bf q}), \\
\label{qsymm}
{\bf J}(p,-{\bf q})&=&-{\bf J}(p,{\bf q}).
\end{eqnarray}
\end{subequations}
A sufficient condition for these two assumptions to hold is that ${\bf
  q} \to - {\bf q}$ corresponds to a spatial reflection of the initial
data.  The assumption (\ref{qsymm}) implies that initial data with
${\bf q}=0$ form a non-spinning black hole, but not that they are
necessarily spherically symmetric.  In the following, for simplicity
of notation, we restrict to axisymmetry, so that the vectors ${\bf q}$
and ${\bf J}$ reduce to their components along the symmetry axis,
which we call simply $q$ and $J$.
  
Generalizing the approximation (\ref{approx_1}) for rotating data we now 
write the evolution in Phase~2 as
\begin{eqnarray}
\label{perturbationregime}
Z(x,\tau)&\simeq& Z_*(x)+\zeta_0(p,q,\tau) Z_0(x)+\zeta_1(p,q,\tau) Z_1(x)
\nonumber \\ &&+
\hbox{(other) decaying modes},
\end{eqnarray}
where $Z_0$ is the single growing spherical mode, and $Z_1$ is either
the single growing $l=1$ axial mode or the least damped such mode.  
In complete analogy to our treatment in Section \ref{sec:theory:ss} the 
amplitudes of the modes are given by
\begin{equation}
\label{zeta0zeta1}
\zeta_0=P(p,q)e^{\lambda_0\tau}, \qquad \zeta_1=Q(p,q)e^{\lambda_1\tau},
\end{equation}
where $\tau$ is again given by (\ref{tau_def}).

The Lyapunov exponent $\lambda_1$ can again be determined from
perturbations of the critical solution \cite{critfluidpert}.
Remarkably, for a perfect fluid with the equation of state
(\ref{eos}), the result can be expressed in closed form as a function
of $\kappa$,
\begin{equation}
\label{lambda1theory}
\lambda_1={1-9\kappa\over3+3\kappa}
\end{equation}
(see Fig.~\ref{Fig:lambdas}). $\lambda_1$ changes sign at $\kappa =
1/9$, marking the transition from $Z_1$ being stable and damped for
$\kappa > 1/9$ to unstable and growing for $\kappa < 1/9$.

From (\ref{symm}) we see that the coefficients $P$ and $Q$ must be
even and odd in $q$, respectively.   Hence the equivalent of
  (\ref{P_expansion_1}) must, to leading order, be
\begin{equation}
Q= C_1 q+\dots 
\end{equation}
for some family-dependent constant $C_1$.  The black hole threshold
within such a two-parameter family is a curve in the $(p,q)$-plane
that is symmetric under $q\to-q$, parameterized by
\begin{equation}
|q|=q_*(p) \qquad \Leftrightarrow \qquad p=p_*(|q|).
\end{equation}
We can fine-tune the initial data to the black hole threshold along
any smooth one-parameter family of initial data that crosses the
threshold, in practice by bisection.  As explained above, we can also
fine-tune to the attracting manifold of the critical solution by
setting $q=0$ and fine-tuning $p$ to the black hole threshold.

 We define the ``reduced
 parameters"
\begin{subequations}
\label{C0C1def}
\begin{eqnarray}
\bp & := & C_0 (p - p_{*0})  \label{C0def} \\
\bq & := & C_1 q\label{C1def}
\end{eqnarray}
\end{subequations}
as shorthands.  Expanding about $\bp=\bq=0$, we approximate
\begin{subequations} \label{PQapproximations}
\begin{eqnarray}
P(\bp, \bq) & = & \bp - K \bq^2 + \ldots, 
\label{Papproximation} \\
Q(\bp, \bq) & = & \bq + \ldots,
\label{Qapproximation}
\end{eqnarray}
\end{subequations}
where $K$ is another family-dependent dimensionless constant. We found
in \cite{BaumgarteGundlach,gb16} that including the $\bq^2$ term is
essential.  Assuming $\bp$ and $\bq^2$ to be of the same order of
smallness, (\ref{PQapproximations}) is a consistent truncation to
leading order.

The parameters $p$ and $q$ stand for any generic parameters of the
initial data that obey (\ref{symm}). In our calculations they will be
represented by $\eta$, which parametrizes the overall fluid density,
and $\Omega$, which controls the rotation rate.  

\subsection{Evolution near the critical solution }

Assume now that $P$ and $Q$ have been chosen small enough so that,
after an initial transition period (Phase~1), $\zeta_0Z_0$ and
$\zeta_1Z_1$ can be treated as linear perturbations of $Z_*$ over some
range $\tau_2\lesssim\tau\lesssim \tau_\sharp$ (Phase~2), before
either perturbation has grown too much and the evolution becomes again
non-perturbative (Phase~3). The starting value $\tau_2$ of $\tau$
depends only weakly on the initial data (through the length scale of
the initial data, rather than the degree of fine-tuning), and its
value does not matter at this point.  ($\tau=0$ itself
  has no particular physical significance, as we have chosen the
  origin of $\tau$ to make $c_M=1$.)
  
From (\ref{zeta0zeta1}), the combination
\begin{equation}
\label{deltazeta}
\delta:=\zeta_1|\zeta_0|^{-\epsilon},
\end{equation}
where
\begin{equation} \label{def_epsilon}
\epsilon:={\lambda_1\over\lambda_0},
\end{equation}
is independent of $\tau$ in Phase~2.  It is therefore a suitable
measure of the relative amplitude of the two modes. In particular,
$\delta$ is given in terms of the initial data $(p,q)$ by
\begin{equation}
\label{deltaQP}
\delta=Q|P|^{-\epsilon}.
\end{equation}
Any function of $\delta$ would of course also be independent of
$\tau$.  The definition (\ref{deltazeta}) is singled out by being odd
in $q$ and to leading order proportional to it. Hence any quantity
that does/does not change sign under a spatial reflection (such as
angular momentum and mass, respectively) must be odd/even in $\delta$.

Many of the qualitative differences between having one or two unstable
modes are a consequence of $\epsilon$ in (\ref{def_epsilon}) becoming
positive when $\lambda_1$ becomes positive, so that $\delta$ in
(\ref{deltaQP}) diverges rather than going to zero as $|P| \to 0$.
For a perfect fluid, however, $\epsilon$ is very small for the entire
range $0\le\kappa\le1/9$, as can be seen from Fig.~\ref{Fig:lambdas},
so that these effects will be difficult to observe numerically.

\subsection{Scaling laws for two unstable modes and universal scaling functions}
\label{subsec:scalinglaws}

We now derive the scaling laws that follow from the existence of a
Phase 2 given by (\ref{perturbationregime}). In contrast to
  \cite{gb16} we allow both $\lambda_0$ and $\lambda_1$ to be
  positive, and in contrast to \cite{scalingfunctions} we explicitly
  treat the two unstable modes on an equal footing.  As $\tau$
increases, one or both of $Z_0$ and $Z_1$ will become nonlinear at
some $\tau=\tau_\sharp$. One possible criterion for this is
\begin{equation}
\label{nonlinearitytmp}
{\zeta_0^2(\tau_\sharp)\over b_0^2}+{\zeta_1^2(\tau_\sharp)\over b_1^2}=1,
\end{equation}
where $b_0$ and $b_1$ are two universal dimensionless constants.
We extend our convention from the non-rotating case to
normalize the mode $Z_0(x)$ so that $b_0=1$, that is, nonlinearity
is given by
\begin{equation}
\label{nonlinearity}
\zeta_0^2(\tau_\sharp)+{\zeta_1^2(\tau_\sharp)\over b_1^2}=1,
\end{equation}
but we retain a nontrivial value of $b_1>0$ in the following.
Adopting this criterion and using (\ref{zeta0zeta1}),
$\tau_\sharp(p,q)$ is given implicitly by
\begin{equation}
\label{xeqn}
P^2e^{2\lambda_0\tau_\sharp} +{Q^2e^{2\lambda_1\tau_\sharp}\over b_1^2}=1.
\end{equation}

\begin{figure}
\ifprintfigures
\includegraphics[width=3.3in]{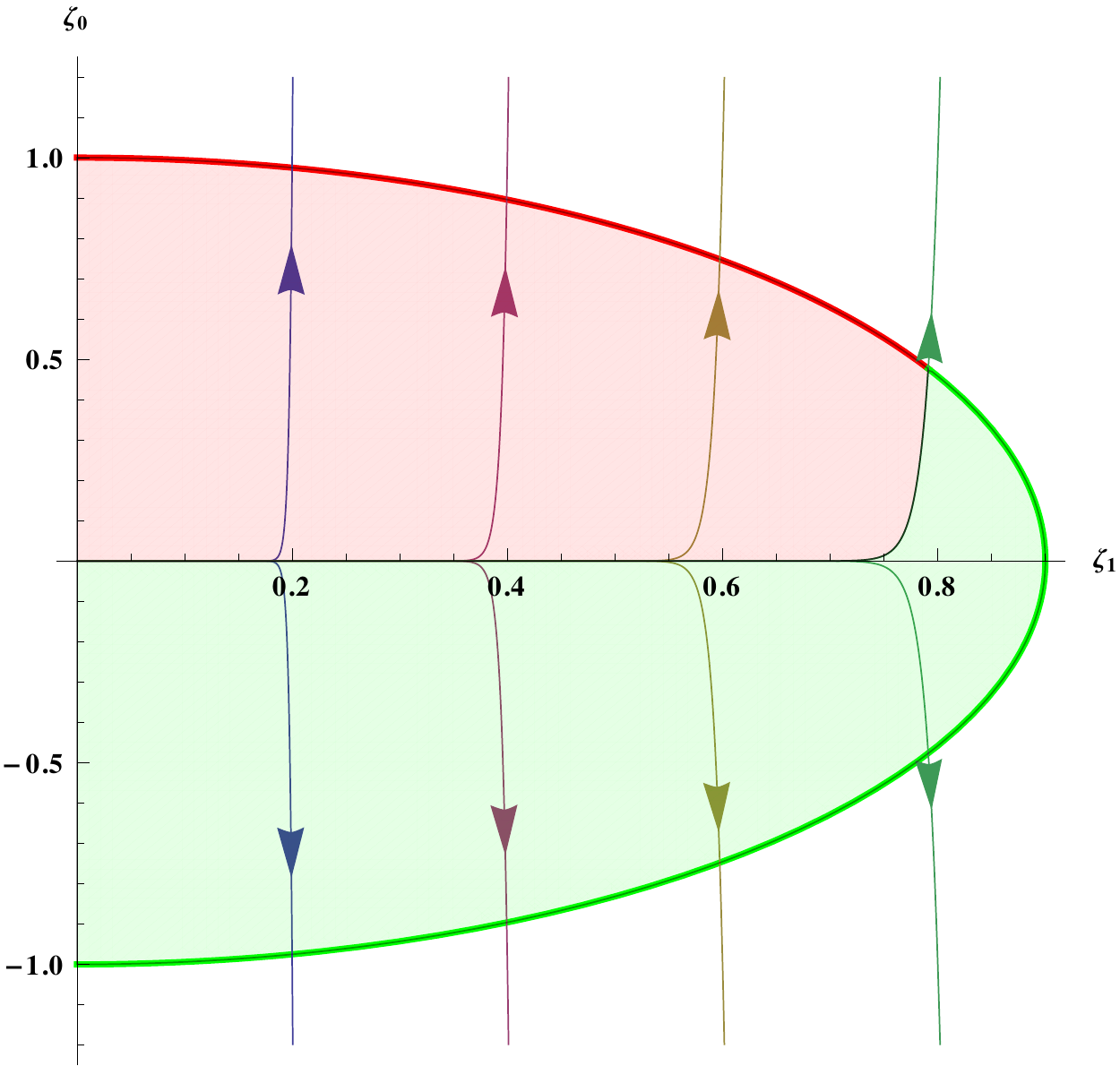}
\fi
\caption{The $\zeta_1\zeta_0$-plane of linear perturbations of
  the critical solution in Phase~2, for $\kappa = 0.08$. Only
  $\zeta_1>0$ is shown. The thin colored lines are lines of constant
  $\delta=0.2,0.4,0.6,0.8$, and mark trajectories of the perturbation
  amplitudes $\zeta_0$ and $\zeta_1$ during Phase~2. These
  trajectories are realistic because they depend only on the known
  parameter $\epsilon\simeq 0.0150$. The thick curve is a {\em
    schematic} representation of the nonlinearity ellipse
  {(\ref{nonlinearity})}. As we do not know the true values of the
  parameters $b_1$ and $(s_*,\delta_*)$, we have {\em arbitrarily} assumed
  $b_1=0.9$ and $(s_*,\delta_*)=(+1,0.8)$. Data on the red segment of
  the nonlinearity ellipse form a black hole in Phase~3, while data on
  the green segment disperse.  Any initial data that go through a
  Phase~2, and hence show critical scaling, must have initial values
  for $(\zeta_1,\zeta_0)$ inside the nonlinearity ellipse. Hence
  evolutions that start Phase~2 in the red shaded region form a black
  hole in Phase~3, while evolutions that start Phase~2 in the green
  shaded region disperse in Phase~3.}
\label{figure:zetaplane008}
\end{figure}

The evolution of the two perturbation modes in the
  $\zeta_1\zeta_0$-plane during Phase 2 is represented in
Fig.~\ref{figure:zetaplane008} for the case $\kappa=0.08$. The thin
lines show evolution trajectories in this plane, i.e.~lines of
constant $\delta$ as defined in (\ref{deltazeta}). As we have
$\lambda_1 > 0$, both modes grow with $\tau$.

Phase~2 ends when the perturbations have grown sufficiently to reach a
point on the nonlinearity ellipse (\ref{nonlinearity}), marked by a
thick line.  The location on the nonlinearity ellipse can be
parametrized by an angle. We can introduce such an angle by
identifying $\zeta_0$ in (\ref{nonlinearity}) with $\cos \alpha$ and
$\zeta_1/b_1$ with $\sin \alpha$. {In
  Fig.~\ref{figure:zetaplane008}, we also schematically show the
  nonlinearity ellipse in $\zeta_1\zeta_0$-plane. The shape of the
  {\em nonlinearity ellipse} is only schematic as we do not currently
  know the true value of $b_1$. Therefore we have simply assumed a value
  that makes for a clear plot and is not in conflict with our
  numerical results. (By contrast, we do have a theoretical
  for the value of $\epsilon$, which alone determines the shape of the
  {\em trajectories}.)}

We may now express the Cauchy data
at the start of Phase 3 of the evolution in two pieces, a scale-invariant part
parameterized by $\alpha$,
\begin{equation}
\label{intermediatedata}
Z=Z_*+\cos{\alpha}\, Z_0+b_1\sin{\alpha}\, Z_1,
\end{equation}
together with an overall length scale given by $e^{-\tau_\sharp}$.
From (\ref{deltazeta}) we see that the angle $\alpha$ is 
related to $\delta$ by
\begin{equation}
\label{alphadelta}
\delta=b_1\sin{\alpha}\left|\cos{\alpha}\right|^{-\epsilon}.
\end{equation}
The range $-\infty<\delta<\infty$ corresponds to the range
$-\pi/2<\alpha<\pi/2$ for $P>0$, and separately to the range
$\pi/2<\alpha<3\pi/2$ for $P<0$. {There is, however, a
  one-to-one correspondence between values of the angle $\alpha$ and
  values of the pair $(s,\delta)$, where $s=\pm 1$ is the sign of
  $P$, or equivalently the sign of $\zeta_0$.}

\begin{figure}
\ifprintfigures
\includegraphics[width=3.3in]{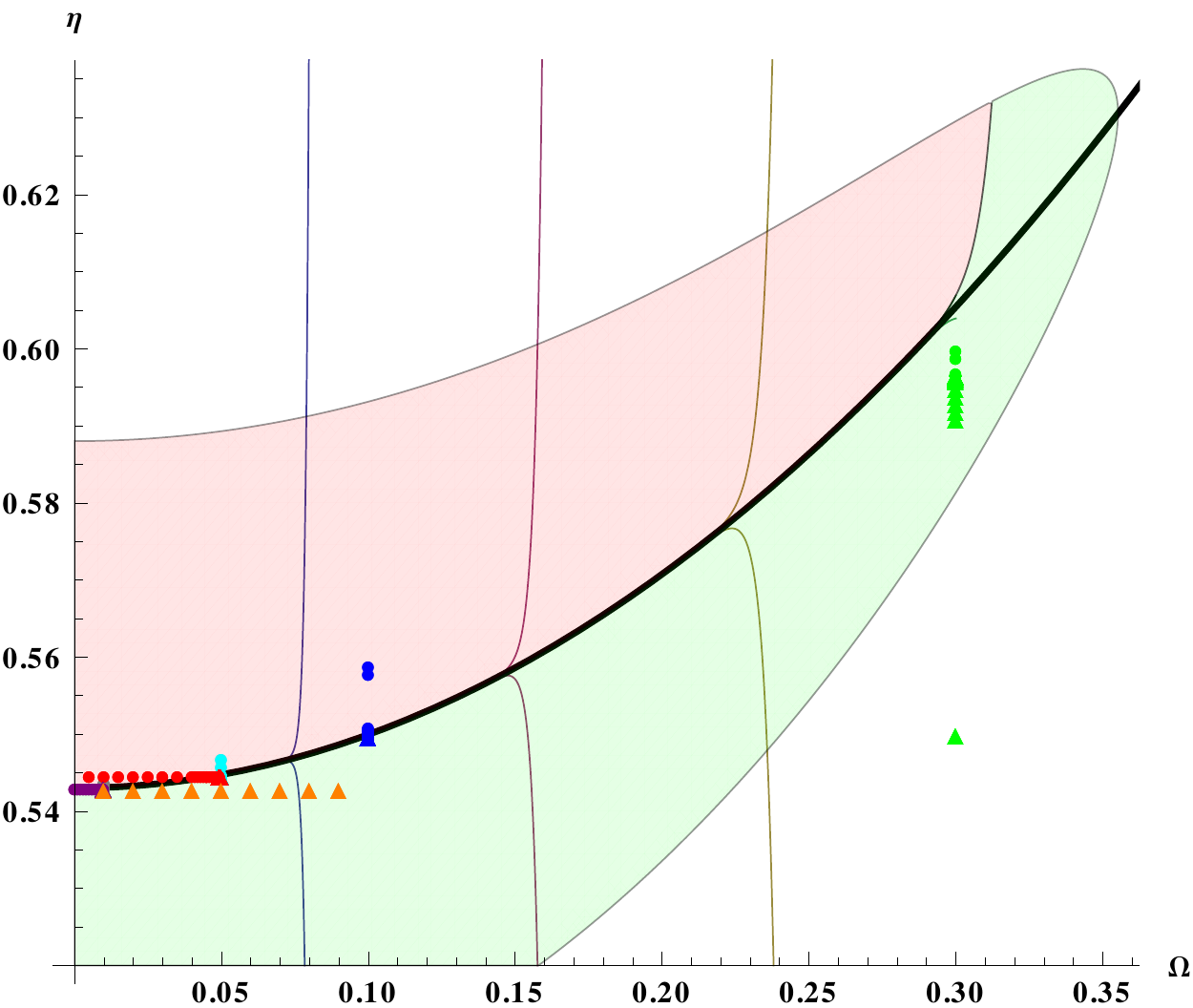}
\fi
\caption{The $\Omega\eta$-plane of initial data for $\kappa =
  0.08$. The coloured dots and triangles represent super and
  subcritical data points, respectively, in our one-parameter families
  of initial data, color-coded as in Table~\ref{table:sequences}. The
  thin coloured curves represent $\delta=0.2,0.4,0.6$, with the same
  color-coding as in Fig.~\ref{figure:zetaplane008}. The black parabola is the
  curve $\delta=\infty$. As discussed in the text, these curves are
  approximately realistic. We also {\em schematically} show which
  initial data {\em would} go through a Phase~2 and form a black hole
  (red shading), or go through a Phase~2 and disperse (green shading),
  for the arbitrarily assumed values $b_1=0.9$, $\delta_*=0.8$ (as in
  Fig.~\ref{figure:zetaplane008}), and $\tau_2=1.9$. As discussed in
  the text, the fact that the black-hole threshold is approximately a
  parabola out to $\Omega=0.3$, shows that $\delta_*$ must be at least
  as large as assumed here, that is $\delta_*>0.8$, and the fact that
  we still see scaling near the black-hole at $\Omega=0.3$ shows that
  the combination $e^{\lambda_1\tau_2}/b_1$ cannot be much smaller
  than assumed here.}
\label{figure:Omegaetaplane008}
\end{figure}

Dimensional analysis now shows that any dimensionless quantity related
to Phase~3 can only depend on the dimensionless angle $\alpha$, but
not on the length scale $e^{-\tau_\sharp}$. In particular, and most
importantly, whether Phase~3 forms a black hole or disperses can
depend only on $\alpha$. In the special case of spherical symmetry, we
have that $\alpha=0$ results in a black hole, while $\alpha=\pi$
results in dispersion. We also know that perturbing spherically
symmetric data with a sufficiently small amount of angular momentum
does not change the final outcome (collapse or dispersion). Hence
there must be a universal constant $0<\alpha_*<\pi$ (depending on the
equation of state parameter $\kappa$) such that Phase~3 forms a black
hole precisely for $-\alpha_*<\alpha<\alpha_*$ (the red part of the
nonlinearity ellipse in Fig.~\ref{figure:zetaplane008}), and disperses
otherwise (the green part of the nonlinearity ellipse.)  This in turn
means that for initial data sufficiently close to both the black hole
threshold and $q=0$ for the evolution to have a Phase~2, the final
outcome (collapse or dispersion) in Phase~3 depends only on
$\alpha(p,q)$ given by (\ref{alphadelta}) with (\ref{deltaQP}).

Furthermore, the dimensionless quantities $J/M^2$ (for supercritical
data) and $\omega_{\rm max}/\sqrt{\rho_{\rm max}}$ (for subcritical
data) can only depend on $\alpha$; that is, there
must exist universal scaling functions $F_{\omega/\sqrt{\rho}}$ and
$F_{J/M^2}$ such that
\begin{subequations}
\label{Falpha}
\begin{eqnarray}
\label{JoM2alpha}
{J\over M^2}&\simeq& F_{J/M^2}(\alpha), \qquad
-\alpha_*<\alpha<\alpha_*, \\
{\omega_{\rm max}\over\sqrt{\rho_{\rm max}}}
&\simeq&F_{\omega/\sqrt{\rho}}(\alpha), \qquad \hbox{otherwise}.
\end{eqnarray}
Here $\rho_{\rm max}$ is the maximum over the entire spacetime of the
central density $\rho_c$, and $\omega_{\rm max}$ is the maximum of
the central angular velocity $\omega_c$ defined below in
Sec.~\ref{sec:diagnostics}.

Finally, any dimensional quantity characterizing Phase~3 must be given
by a suitable power of the length scale $e^{-\tau_\sharp}$, times a
universal scaling function of ${\alpha}$. For supercritical evolutions
these quantities include the black hole mass $M$ and angular momentum
$J$, while for subcritical evolutions they include the maximum values
taken by the central fluid density $\rho_c$ and the central angular
velocity $\omega_c$. Hence we must have
\begin{eqnarray}
M&\simeq& e^{-\tau_\sharp}F_M({\alpha}), \\
J&\simeq& e^{-2\tau_\sharp}F_J({\alpha}), \\
\rho_{\rm max}&\simeq& e^{2\tau_\sharp}F_\rho({\alpha}), \\
\omega_{\rm max}&\simeq& e^{\tau_\sharp}F_\omega({\alpha}),
\label{omegamaxalpha}
\end{eqnarray}
\end{subequations}
where obviously $F_M$ and $F_J$ are defined only for
$-\alpha_*<\alpha<\alpha_*$ and $F_\rho$ and $F_\omega$ for the
complementary range of $\alpha$, and where $F_{J/M^2}=F_M/F_J^2$ and
$F_{\omega/\sqrt{\rho}}=F_\omega/\sqrt{F_\rho}$.  

{
We could express $\alpha_*$ in terms of the pair
$(s_*,\delta_*)$. Similarly, we could replace the angle $\alpha$ as
the argument of the universal scaling functions $F$ by the pair
$(s,\delta)$, but this would then require a pair of scaling functions
$F_\pm(\delta)$. Another disadvantage of using $(s,\delta)$ is that,
at $P=0$, $\delta$ is not a smooth parameter. By contrast, $\alpha$
can be smoothly continued across $P=0$, with $\alpha\pm\pi/2 \simeq
b_1^{-1/\epsilon} P$, and across $Q=0$, with $\alpha\simeq b_1Q$ and
$\pi-\alpha\simeq b_1Q$.}

$F_M$ and $F_\rho$ must be even functions of $\alpha$ (or $\delta$),
while $F\omega$, $F_J$, $F_{\omega/\sqrt{\rho}}$ and $F_{J/M^2}$ are
odd. Hence at $\alpha\simeq 0$ ($\delta\simeq 0$ with $P>0$), and
using our previous conventions, to leading order we must have
\begin{subequations}
\label{Fleading}
\begin{eqnarray}
\label{FMleading}
F_M&\simeq& 1, \\
F_J&\simeq& \delta \simeq b_1\alpha, \\
F_{J/M^2}&\simeq& \delta \simeq b_1\alpha, 
\end{eqnarray}
while at $\alpha\simeq\pi$ ($\delta\simeq 0$ with $P<0$), we must have
\begin{eqnarray}
F_\rho &\simeq& c_\rho, \\
F_\omega &\simeq& c_\omega \delta \simeq c_\omega b_1(\pi-\alpha), \\
F_{\omega/\sqrt{\rho}}&\simeq& {c_\omega\over\sqrt{c_\rho}} \delta 
\simeq {c_\omega\over\sqrt{c_\rho}} b_1(\pi-\alpha),
\label{Fomegaosqrtrholeading}
\end{eqnarray}
\end{subequations}
where we have defined the shorthands $c_\rho:=F_\rho(\delta=0)$ and
$c_\omega:=dF_\omega/d\delta(\delta=0)$.  (In contrast to
Fig.~\ref{figure:zetaplane008}, these curves are {\em} not
trajectories, as the time evolution of our initial data does not stay
within our family of initial data.)
Fig.~\ref{figure:Omegaetaplane008} shows {theoretical} curves of
constant $\delta$ in the $\Omega\eta$-plane of initial data for the
two-parameter family of initial data described below, for $\kappa =
0.08$. Note that, in contrast to the $\zeta_1\zeta_0$-plane in
Fig.~\ref{figure:zetaplane008}, these curves are not trajectories, as
the time evolution of the initial data does not remain within our
two-parameter family of initial data. {To compute $\delta$, we
  have used the definition (\ref{deltaQP}) of $\delta$ in terms of $P$
  and $Q$, together, with the leading-order approximations
  (\ref{PQapproximations}) of $P$ and $Q$ in terms of $\Omega$ and
  $\eta$, the known values of $\lambda_0$ and $\lambda_1$, and the
  numerically fitted values of the family-dependent parameters
  $\eta_{*0}$, $C_0$, $C_1$ and $K$. Hence theses curves of constant
  $\delta$ are realistic approximations. In the same figure,} we also show
the one-parameter (sub)families of initial data by which we have
explored our two-parameter family.  {This juxtaposition gives
  rise to constraints on $\delta_*$ and $b_1$ as follows.

First, the observed black-hole threshold in the $\Omega\eta$-plane is
to good approximation a parabola out to $\Omega=0.3$. This means it is
indistinguishable from what our prediction for the black-hole
threshold would be for $\delta_*=\infty$. To make this more
quantitative, we have plotted what the predicted black-hole threshold
{\em would} be for $(s_*,\delta_*)=(+1,0.8)$ (the same arbitrary assumption
we already made in Fig.~\ref{figure:zetaplane008}). This
hypothetical threshold turns up sharply just beyond $\Omega=0.3$, and
so is marginally compatible with our numerical observations, giving a
hard constraint of $\delta_*\ge 0.8$ on $\delta_*$. However, the specific
value $\delta_*=0.8$ we have chosen Figs.~\ref{figure:zetaplane008}
and \ref{figure:Omegaetaplane008} is for schematic illustration only.}

Second, the region {in the $\Omega\eta$-plane} in which scaling
is observed provides a (much less clear-cut) constraint on $b_1$.  As
before, we let $\tau_2$ be the value of $\tau$ at the beginning of
Phase~2. This value depends on the family of initial data, and within
the family should depend smoothly on the initial data, and in
particular varying little over data near the black-hole
threshold. Then the condition that the evolution at $\tau_2$ is still
within the nonlinearity ellipse is that
\begin{equation}
\label{tau2constraint}
P^2e^{2\lambda_0\tau_2} +{Q^2e^{2\lambda_1\tau_2}\over b_1^2}<1.
\end{equation}
Hence data with $(P,Q)$ that obey this condition will admit a Phase~2
and hence will be in the critical scaling regime. For given values of
$\tau_2$ and $b_1$, this defines an ellipse in the $QP$-plane, and
hence a deformed ellipse in the $\Omega\eta$-plane. 

{In Fig.~\ref{figure:Omegaetaplane008} we have chosen arbitrary
  values $\tau_2=1.9$ and $b_1=0.9$ (the latter the same as in
  Fig.~\ref{figure:zetaplane008}) that show the shape of this region
  clearly, while still being compatible with our observations of
  scaling at $\Omega=0.3$. From this figure we see that the
  combination $e^{\lambda_1\tau_2}/b_1$ of $\tau_2$ and $b_1$ cannot
  be much smaller than we have assumed here in order to make the
  shaded region extend all the way to $\Omega=0.3$.

We stress again that the values of $(s_*,\delta_*)$, $b_1$ and
$\tau_2$ in Figs.~\ref{figure:zetaplane008} and
\ref{figure:Omegaetaplane008}, while compatible with our numerical
observations, have been chosen primarily to give clear schematic plots
of the shaded regions.}


\subsection{Review of the case of one unstable mode}


\begin{figure}
\ifprintfigures
\includegraphics[width=3in]{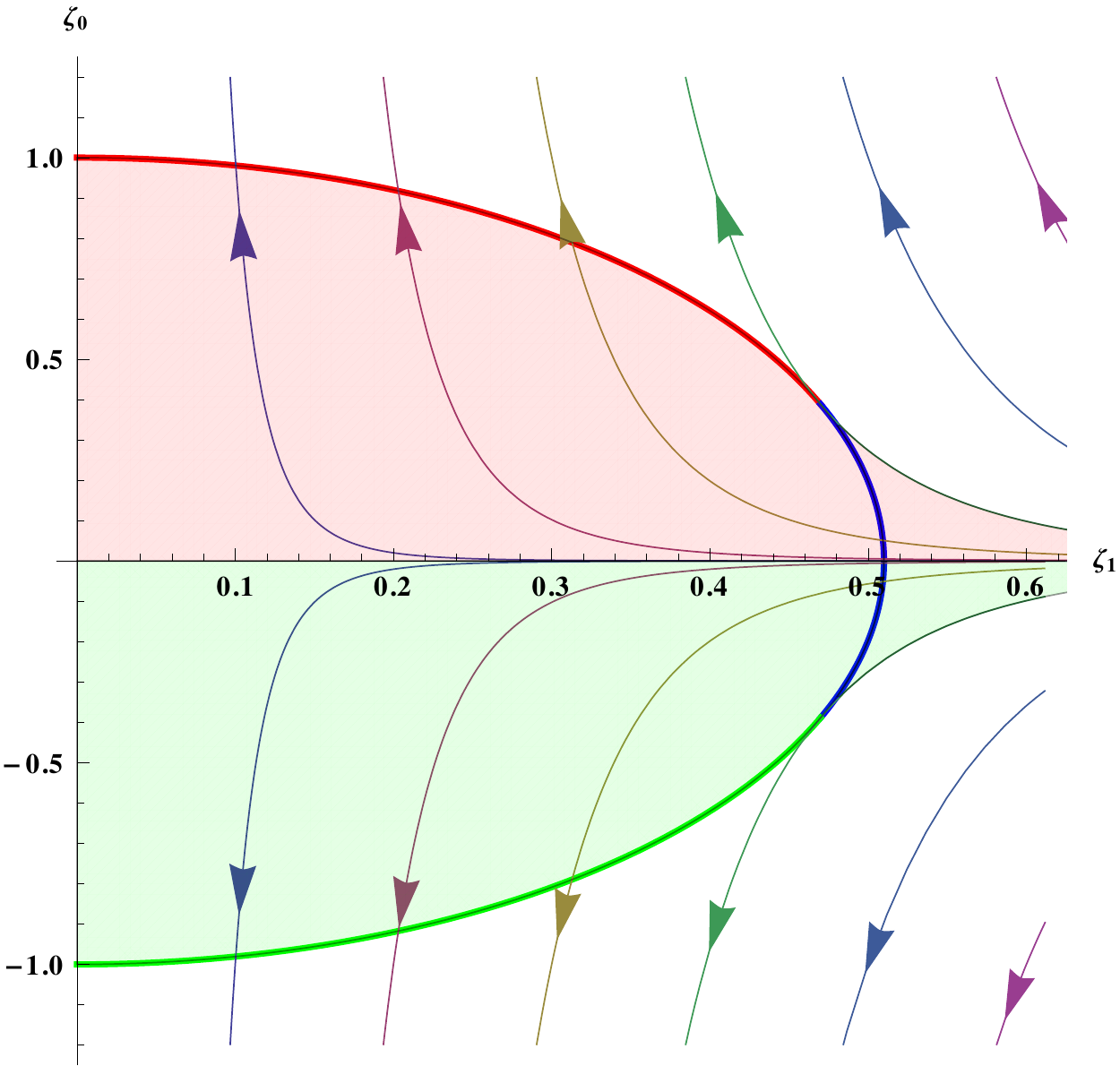}
\fi
\caption{{The $\zeta_1\zeta_0$-plane of linear perturbations of
    the critical solution in Phase~2, for $\kappa = 1/3$. The thin
    colored lines are lines of constant $\delta=0.1,0.2,\dots,0.6$,
    and mark trajectories of the perturbation amplitudes $\zeta_0$ and
    $\zeta_1$ during Phase~2. We have used the known value
    $\epsilon\simeq-0.178$, and hence these curves are realistic. The
    thick curve is a {\em schematic} representation of the
    nonlinearity ellipse {(\ref{nonlinearity})}. As we do not know the
    true value of $b_1$, we have arbitrarily assumed that $b_1=0.51$,
    giving $\delta_{\rm max}\simeq 0.4$.} Data on the red segment of
  the nonlinearity ellipse form a black hole in Phase~3, while data on
  the green segment disperse.  Trajectories {\em enter} the
  nonlinearity ellipse across the blue segment. To show critical
  scaling, a time evolution must go through a Phase~2, that is, there
  must be an interval where it can be represented by a trajectory
  inside the nonlinearity ellipse. {Hence evolutions that start
    in the red shaded region show scaling during Phase~2 and form a
    black hole in Phase~3, while evolutions that start in the green
    shaded region show scaling during Phase~2 and disperse in
    Phase~3. Note that the trajectories, and hence the boundary of the
    shaded regions, are only schematic outside the ellipse}, as the
    evolution is non-perturbative there.}
\label{figure:zetaplane1o3}
\end{figure}

At this point, it may be instructive to revisit the case $\epsilon<0$ of a
single unstable mode in the light of our notation for $\epsilon>0$. 
There are several important qualitative differences.

First, Eq.~(\ref{xeqn}) shows that for $\lambda_1<0$, $\tau_\sharp \to
\infty$ as $P \to 0$ for any value of $Q$. This means that we can
achieve arbitrarily small length scales $e^{-\tau_\sharp}$ with
sufficient fine-tuning of $P$ alone.
 
Second, (\ref{deltaQP}) with $\epsilon<0$ implies that
$\delta \to 0$ as the black hole threshold is approached and the
approximations (\ref{Fleading}) become
increasingly accurate. 

Third, as illustrated by Fig.~\ref{figure:zetaplane1o3} for the case
$\kappa=1/3$, there is a trajectory in each quadrant of the
$\zeta_1\zeta_0$-plane that just touches the nonlinearity ellipse
$\zeta_0^2+(\zeta_1/b_1)^2=1$. A simple calculation shows that in the
upper right quadrant the touching point is at $\alpha=\alpha_{\rm
  max}$ given by 
\begin{equation}
\label{alphamax}
\alpha_{\rm max}:={\rm arccot}\sqrt{-\epsilon}, 
\end{equation}
and
similarly in the other three quadrants.  (For given $b_1$,
$\alpha_{\rm max}$ defines an equivalent $\delta_{\rm max}$.) Hence in
the range $\alpha_{\rm max}<|\alpha|<\pi-\alpha_{\rm max}$
trajectories enter the nonlinearity ellipse, rather than leaving it.
Therefore, $\alpha_*$ is not defined for $\epsilon < 0$. Rather, black
holes form if and only if $\zeta_0>0$, that is $P>0$, as illustrated
in Fig.~\ref{figure:zetaplane1o3}. {
The nonlinearity ellipse in this figure is only schematic, as we do
not know the true value of $b_1$. For plotting, we have made the
arbitrary assumption $b_1=0.51$, which, from (\ref{alphamax}) and
(\ref{alphadelta}), yields $\delta_{\rm max}\simeq
0.4$. As we observed values of $J/M^2$ up to $0.4$ in the scaling
regime, and we approximate $J/M^2\simeq \delta$ in the case of a
single unstable mode, our assumed value of $b_1$ is just compatible
with our observations, and we must have the constraint $b_1\ge 0.51$.}

\begin{figure}
\ifprintfigures
\includegraphics[width=3.3in]{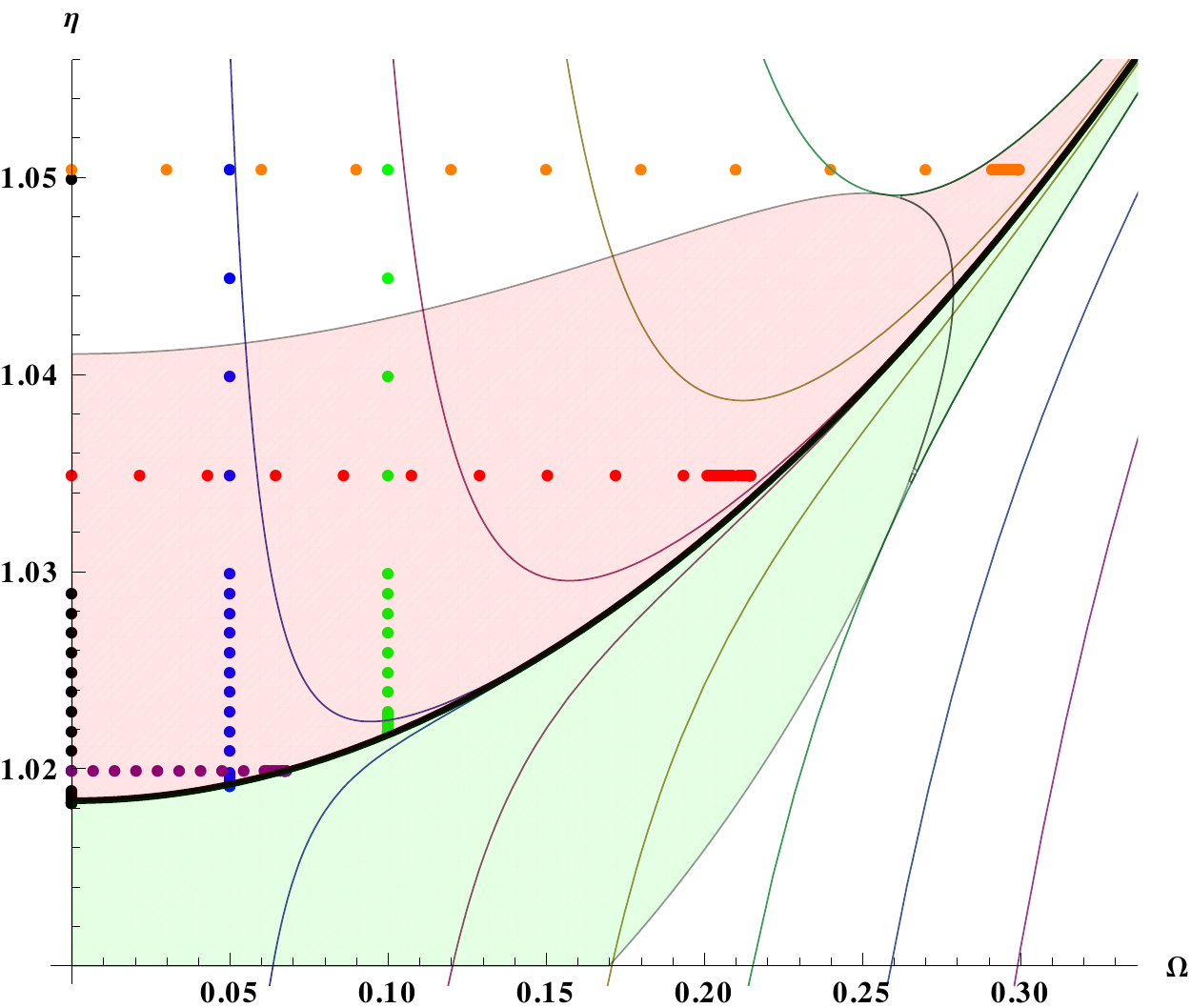}
\fi
\caption{{The $\Omega\eta$-plane of initial data for $\kappa =
    1/3$.} We show the curves $\delta=0.1,0.2,\dots,0.6$, with the
  same color-coding as in Fig.~\ref{figure:zetaplane1o3}.  We also
  show the one-parameter families of initial data used in \cite{gb16},
  color-coded as in Fig.~2 of~\cite{gb16}. (All these data are
  supercritical, and hence represented by dots.)  The black hole
  threshold is at $P=0$, that is at $\delta=0$ (thick black
  line). {As discussed in the text,} these curves are
  approximately realistic. {We also {\em schematically} show the region
  of initial data space in which we {\em would} expect to see scaling
  under the arbitrary assumptions that $b_1=0.1$ (as in
  Fig.~\ref{figure:zetaplane1o3}) and $\tau_2=1.8$.} Any smaller value
  of $b_1$ would give a smaller $\delta_{\rm max}$ and would therefore
  be ruled out by the observation that we observe values of
  $\delta\simeq J/M^2$
  up to $0.4$ in the scaling regime.}
\label{figure:Omegaetaplane1o3}
\end{figure}

Trajectories with $|\delta|>\delta_{\rm max}$ miss the nonlinearity
ellipse altogether, meaning that there is no Phase~2 and hence no
critical scaling. However, if we extrapolate the linear perturbation
picture beyond its assumed region of validity to the entire
$\zeta_1\zeta_0$-plane, then evolutions with arbitrarily large
$|\zeta_1|$ will eventually enter the nonlinearity ellipse and go
through a Phase~2, provided that $|\zeta_0|$ is sufficiently small,
namely that $|\delta|<\delta_{\rm max}$. It is natural to assume that
the linear perturbation picture is still a qualitatively correct
description of the full nonlinear dynamics. This is assumed in
Figs.~\ref{figure:zetaplane1o3} and \ref{figure:Omegaetaplane1o3} in
extending the shaded regions beyond the original ellipses. 

If the attracting manifold of the critical solution were identical
with the entire black hole threshold in the space of initial data, we
would see critical scaling in the time evolution of {\em all} data
that are sufficiently close to the black hole threshold, even though
very far from initial data for the critical solution.

There is good numerical evidence that in spherical symmetry, for
various types of matter, the attracting manifold of the critical
solution is indeed the entire black-hole threshold, including
initial data which are very far from the critical solution. Beyond
spherical symmetry, this assumption can be true only in some local
sense. To construct an obvious counterexample, a two-parameter family
of axisymmetric initial data could consist of two blobs of fluids
(rotating for $q\ne 0$) well separated along the symmetry axis, each
of which might show critical phenomena, but separately at different
threshold values of $p$. However, in a (potentially large)
neighborhood of spherically symmetric initial data we expect the
perturbative phase space picture to be a qualitatively correct model
of the full phase space picture, that is, as far as the attracting
manifold of the critical solution extends as a smooth submanifold of
the space of initial data, its codimension must be given by the number
of its unstable perturbation modes of the critical solution. Further
away, it could end or change dimension in an unsmooth manner, for
example in a caustic.

For the non-rotating scalar field this was shown in \cite {ChoHLP03}
(although their results also seem to indicate a second, non-spherical,
unstable mode) and for the non-rotating radiation fluid in
\cite{BaumgarteMontero}. We leave to future work the question how far
the attracting manifold of the spherical critical solution extends
into the space of initial data also for rotating perfect fluids (for
the equations of state $1/9<\kappa\lesssim 0.49$ where the critical
solution has a single unstable mode).

\subsection{Breakdown of scaling at sufficient fine-tuning}

Returning now to the case of two unstable modes, by contrast, we have
$\lambda_1 > 0$.  Now, from Eq.~(\ref{xeqn}), $\tau_\sharp \to \infty$
requires that $P$ and $Q$ are {\rm both} fine-tuned to zero. This is
consistent with the observation that, in the presence of two unstable
modes, the attracting manifold of the critical solution has
codimension two. Conversely, for any $Q > 0$, the length scale will
remain finite even when $P$ is fine-tuned to zero.  Eq.~(\ref{xeqn})
gives $\tau_\sharp(P,Q)$ only in implicit form, but we can obtain a
rough explicit approximation by assuming that one of the two terms on
the left-hand side of (\ref{xeqn}) always dominates, giving
\begin{equation}
\left(e^{-\tau_\sharp}\right)\simeq \max\left(
|P|^{1\over\lambda_0},|Q/b_1|^{1\over\lambda_1}\right).
\end{equation}
In practice, we cross the black hole threshold on {one}-parameter families
of initial data on which $Q$ remains finite but on which $P$ changes
sign. For such families we then have 
\begin{equation}
\label{minscale}
\left(e^{-\tau_\sharp}\right)_{\rm min}\simeq |Q/b_1|^{1\over\lambda_1}.
\end{equation}
Equivalently, along such families we expect a breakdown of scaling when
$P$ is small enough such that the second term on the left-hand side of
(\ref{xeqn}) dominates, that is approximately for 
\begin{equation}
\label{scalingbreakdown}
|P|<|Q/b_1|^{1\over \epsilon}.
\end{equation}

Note, however, that the existence of a {minimum of the 
  overall length scale $e^{-\tau_\sharp}$} does not necessarily imply
that $M$, $J$, $\rho_{\rm max}$ and $\omega_{\rm max}$ are finite on
the black hole threshold. Our theoretical model does not tell us the
behaviour of the scaling functions $F(\alpha)$ at $\alpha_*$. Any or
all of them could vanish there, or be finite.  If they all vanish,
this may mask the formal existence of the {minimum of the
  overall length scale $e^{-\tau_\sharp}$} given by (\ref{minscale}),
although scaling would still break down, unless the scaling functions
also approached zero as a power of $|\alpha|-\alpha_*$.

\subsection{Leading-order power laws at the black hole threshold}

Returning to our discussion of the case of two unstable modes, it is
possible that the black hole threshold occurs at $P = 0$,
i.e.~$\delta_* = \infty$ and $\alpha_* = \pi/2$.  This is not a given,
however, and in this section we will discuss the form of leading-order
power laws at the black hole threshold without making this assumption,
in contrast to \cite{scalingfunctions}.

For $\alpha_*\ne \pi/2$, or equivalently $\delta_*\ne \infty$, a
generic one-parameter family of initial data crosses the black hole
threshold at nonzero $Q$ and $P$ (giving a finite
$\delta=\delta_*$). However, because $\epsilon$ is so small, this will
happen at a very small value of $P$, and it will be difficult to
distinguish the point on a one-parameter family of initial data where
it crosses the black hole-threshold at $|P|=|Q/\delta_*|^{1/\epsilon}$
from the point where $P=0$ occurs.  A different way of saying this is
that near the black hole threshold $\alpha$ changes very quickly along
any generic one-parameter family of initial data, so that $\alpha_*$
[or equivalently $(s_*,\delta_*)$] is difficult to determine from the
location of the black hole threshold, see
Fig.~\ref{figure:Omegaetaplane008}.

In principle {it would be possible to} determine the universal
constants $\alpha_*$ and $b_1$ and the universal scaling functions
$F(\alpha)$ directly by evolving the one-parameter family of initial
data (\ref{intermediatedata}) at the beginning of Phase 3.  This,
however, requires knowledge of the functions $Z_*$, $Z_0$ and $Z_1$
in a gauge adapted to our code, which we currently do not
have.  We instead evolve fine-tuned generic initial data,
which, we assume, ultimately evolve into
(\ref{intermediatedata}). This means that in testing our theoretical
model against numerical data such as $M(p,q)$, we have to fit
$\alpha_*$, $b_1$ and the $F(\alpha)$, as well as (necessarily) the
family-dependent functions $P(p,q)$ and $Q(p,q)$.

As a way around our ignorance of $\alpha_*$, we slightly recast the
theory in a way that relates more directly to observation, and in
particular uses the fact that, in contrast to $P$ and $Q$, the black
hole threshold $p=p_*(|q|)$ is directly observable. We note that the
scaling laws (\ref{Falpha}) do not change their
form if we replace $\tau_\sharp$ as a measure of overall scale with
\begin{equation}
\tau_\flat:=\tau_\sharp+y(\alpha).
\end{equation}
In particular, we choose $y(\alpha)$ to be defined implicitly by
\begin{equation}
\label{Ptilde}
e^{-\lambda_0\tau_\flat}:=\tilde P:=P-s_*
\left(|Q|\over\delta_*\right)^{1\over\epsilon}
{=P\left[1-ss_*\left(|\delta|\over\delta_*\right)^{1\over\epsilon}\right]}.
\end{equation}
As defined before, $s$ is the sign of $P$ and $(s_*,\delta_*)$ is an
alternative parameterisation of $\alpha_*$, so that $s_*=1$ for
$0<\alpha_*<\pi/2$ and $s_*=-1$ for $\pi/2<\alpha_*<\pi$. At the same
time, instead of either $\alpha$ or $\delta$ we use
\begin{equation}
\label{deltatilde}
\tilde\delta:=Q|\tilde P|^{-\epsilon}
\end{equation}
as the argument of our scaling functions. This guarantees that $\tilde
P=0$ and $\tilde \delta=\infty$ at the black-hole threshold
  $(s,\delta)=(s_*,\delta_*)$. In the limit $\delta_*=\infty$ (or
{$\alpha_*=\pi/2$}) we recover $\tilde P=P$ and
$\tilde\delta=\delta$. It is easy to derive an explicit expression for
$\tilde\delta(\delta,s_*,\delta_*)$ and an algebraic equation giving
$y(\alpha)$ implicitly, but their exact forms do not matter here.

We may therefore write our scaling formulas as
\begin{subequations} 
\label{scaling}
\begin{eqnarray}
M&\simeq& e^{-\tau_\flat} \tilde F_M( \tilde \delta), \\
J&\simeq& e^{-2\tau_\flat}  \tilde F_J( \tilde \delta) , \\
\rho_{\rm max}&\simeq& e^{2\tau_\flat}  \tilde F_\rho( \tilde \delta), \\
\omega_{\rm max}&\simeq& e^{\tau_\flat}  \tilde F_\omega( \tilde \delta) .
\end{eqnarray}
\end{subequations}
For $|\delta|\ll \delta_*$, Eq.~(\ref{Ptilde}) gives $\tilde P\simeq P$, and
  hence $\tau_\sharp\simeq \tau_\flat$ and $\tilde\delta\simeq
\delta$. The leading-order expressions for the $\tilde
F(\tilde\delta)$ at small $\tilde \delta$ are therefore the same as
for the original scaling functions at small $\delta$,
Eqs.~(\ref{Fleading}).

Given a two-parameter family of initial data, the values of the
family-dependent constants $C_0$ and $C_1$ are then fixed by imposing
the conventions $\tilde F_M=1$ and $d\tilde F_J/d\tilde \delta=1$ at
$\tilde\delta=0$. The sign of $C_0$ is fixed by the convention that at $q=0$
black holes form for $P>0$.

For any {one}-parameter family of initial data that crosses the black hole
threshold at $\tilde P=0$ but $Q\ne 0$, at the point $(p_*,q_*)$, we
have, from Eqs.~(\ref{C0C1def}) and (\ref{PQapproximations}),
\begin{eqnarray}
\label{Ptildeapprox}
\tilde P&\simeq& C_0(p-p_*)+KC_1^2(q_*^2-q^2) \nonumber \\ 
&\simeq& C_0(p-p_*)+2KC_1^2q_*(q_*-q),
\end{eqnarray}
that is, to leading (linear) order $\tilde P$ is simply the distance to the
black hole threshold, a directly observed quantity. 

In theory, we pay a price for this convenient re-parameterization. For
example, if $M$ is actually finite at the black hole threshold, $\tilde
F_M$ must diverge there to compensate the fact that $e^{-\tau_\flat}$
vanishes there. However, we do not see any indication of a
  breakdown of scaling in any of $M$, $J$, $\rho_{\rm max}$ or
$\omega_{\rm max}$ at the level of numerical fine-tuning we can
achieve, and so the parameterization of our initial data in terms of
$\tilde P$ and $\tilde\delta$ has no downside in our applications.

We note that the expressions (\ref{scaling}) do not form power-law
scaling laws yet. However, for $Q$ not too large and $\tilde P$ not
too small, we will have small $\tilde\delta$ (or $\delta$), and so we
can use (\ref{Fleading}) to obtain the leading-order power laws
\begin{subequations} \label{powerlaws} 
 \begin{eqnarray}
 M & \simeq &  \tilde P^{\gamma_M}, 
 	\label{Mpowerlaw} \\
{J\over Q}  & \simeq & \tilde P^{\gamma_J}, 
 	\label{Jpowerlaw} \\
 \rho_{\rm max} & \simeq & c_\rho  (-\tilde P)^{\gamma_\rho}, 
 	\label{rhopowerlaw} \\
 {\omega_{\rm max} \over Q}
 & \simeq & c_\omega (-\tilde P)^{\gamma_\omega}, \label{omegapowerlaw}
 \end{eqnarray}
 \end{subequations}
 where we have identified the critical exponents
\begin{subequations} \label{crit_exponents}
\begin{eqnarray}
\gamma_M & =  &  {1 \over \lambda_0},\\
\gamma_J & = & {2 - \lambda_1\over \lambda_0}, \\
\gamma_\rho & =  &  -{2  \over \lambda_0}, \\
\gamma_\omega & =  &  - {1 + \lambda_1\over \lambda_0}.
\end{eqnarray}
\end{subequations}
(We assume again the convention that a black hole forms for $\tilde P>0$.)
Eqs.~(\ref{powerlaws}) generalize the scaling law
(\ref{choptuikscaling}) to rotating configurations. 
We also see that, in this limit, the dimensionless quantities become
\begin{subequations}
\begin{eqnarray}
\label{JM2delta}
{J \over M^2} &\simeq& Q \tilde P^{-\epsilon}, \\
{\omega_{\rm max} \over \sqrt{\rho_{\rm max}}} &\simeq&
{c_\omega \over \sqrt{c_\rho}} Q (-\tilde P)^{-\epsilon}.
\end{eqnarray}
\end{subequations}
 
\section{Numerics}

\subsection{Numerical Code}

We employ a numerical code that solves Einstein's equations, expressed
in the BSSN formalism
\cite{Nakamura,ShibataNakamura,BaumgarteShapiro}, in spherical polar
coordinates (details of our implementation can be found in
\cite{Baumgarteetal13,Baumgarteetal15}).  The code makes no symmetry
assumptions, and handles the coordinate singularities using a
reference-metric formalism \cite{Brown,Gourgoulhon} together with a
proper rescaling of all tensorial quantities.

For the simulations presented in this paper we adopt axisymmetry,
simply by setting all derivatives with respect to the azimuthal
angular variable $\varphi$ to zero, and by using the smallest possible
number of grid points in the $\varphi$-direction.  We also impose
equatorial symmetry between the two hemispheres, and resolve one
hemisphere with a very modest number of $N_\theta = 12$ angular grid
points.  Since all functions depend on angle much more weakly than on
radius, we expect that the resulting numerical error is still
relatively small (see Fig.~7 in \cite{BaumgarteMontero} for a
demonstration).  We use $N_r = 312$ radial grid points, which are
allocated logarithmically (see Appendix A in \cite{BaumgarteMontero}),
so that the ratio between the size of the innermost and the outermost
grid cells is about 0.0021.  During the evolution we compute a typical
length-scale of the solution at the origin from $l = (\rho /
\partial^2_r \rho)^{1/2}$, and compare this length-scale with the size
$\Delta_r$ of the innermost radial grid-cell.  When $\Delta_r / l$
exceeds a certain tolerance (typically set to {0.05}) we re-grid, 
meaning that we move the outer boundary to a
smaller location, and interpolate all grid-functions to a new grid
with the same grid number and logarithmic cell distribution.  
We re-grid up to 10 times, moving the outer boundary from 72 (in our code
units) to a minimum value of 12 in equal fractions.  This ensures that the
center of the simulations remains outside of the domain of dependence of the
outer boundary for sufficiently long for a black hole to settle down
(see Fig.~\ref{figure:horizons} below).

As in \cite{BaumgarteGundlach,gb16} we carry out our simulations with
a 1+log slicing condition for the lapse $\alpha$ \cite{Bonaetal} as
well as the version of the Gamma-driver condition
\cite{Alcubierreetal} presented in \cite{Thierfelderetal} for the
shift vector $\beta^i$.

\subsection{Initial Data}
\label{sec:indata}

We adopt the same two-parameter family of initial data as in
\cite{BaumgarteGundlach,gb16}, except that we now allow for a general
$\kappa$ in the equation of state (\ref{eos}).  Specifically, we set
up a momentarily static spherically symmetric fluid ball with a
Gaussian density distribution, centered on the origin, and
parametrized by the central density $\eta$ (see Eq.~(6) in
\cite{BaumgarteGundlach}).  We then endow this fluid with an angular
velocity parametrized by $\Omega$ (see Eq.~(7) in
\cite{BaumgarteGundlach}) and solve the Hamiltonian and momentum
constraints iteratively until the solution has converged to a desired
tolerance.  The parameters $\eta$ and $\Omega$ are our specific
instantiations of the parameters $p$ and $q$ used in Section
\ref{sec:theory}. {Accordingly} the amplitude of $Z_1$ must vanish for
$\Omega=0$.

We also set the shift to zero initially, and choose a ``pre-collapsed"
lapse $\alpha = \psi^{-2}$, where $\psi$ is the conformal factor.

\subsection{Diagnostics}
\label{sec:diagnostics}

We monitor a number of different quantities during our evolutions.

For supercritical evolutions we locate apparent horizons
\cite{ShibataUryu} and measure their irreducible mass $M_{\rm irr}$
and angular momentum $J$ (see \cite{Dreyeretal}).  After a horizon is
first formed, it grows for some time, as the newly formed black hole
accretes more mass, but ultimately settles down to an equilibrium.
From these equilibrium values of $M_{\rm irr}$ and $J$ we determine
the Kerr mass
\begin{equation} \label{MKerr}
M = M_{\rm irr} \left( 1 + \frac{1}{4} \left(\frac{J}{M_{\rm irr}} \right)^2 \right)^{1/2}.
\end{equation}

\begin{figure}
\ifprintfigures
\includegraphics[width=3.7in]{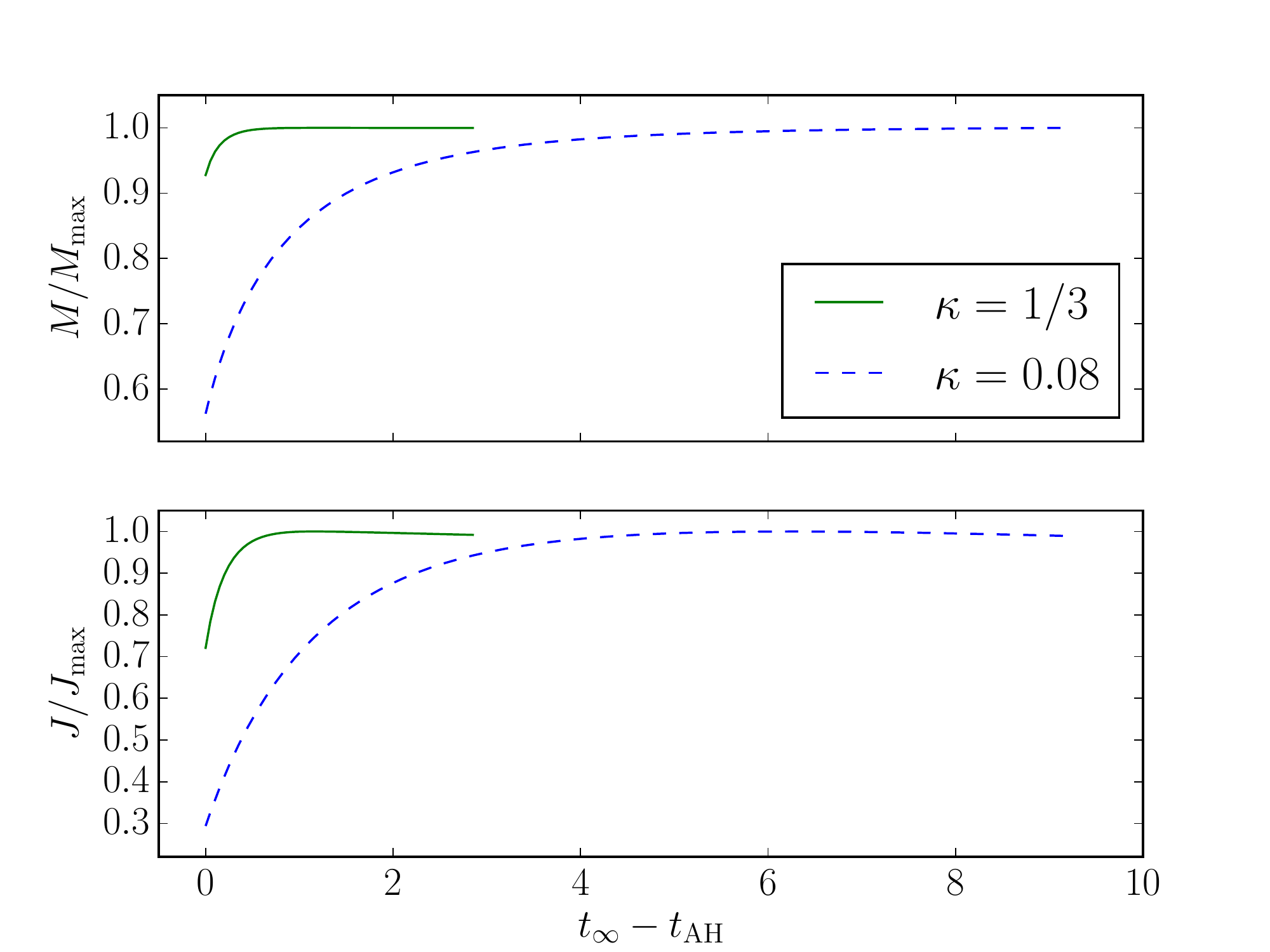}
\fi
\caption{The horizon mass $M$ and angular momentum $J$, divided by
  their maximum values, as a function of coordinate time, which agrees
  with the proper time $t_\infty$ as measured by a static observer at
  infinity.  We denote with $t_{\rm AH}$ the time at which a horizon
is first detected.  We show examples for $\kappa = 1/3$ and $\kappa =
0.08$, both for $\Omega = 0.3$, and both for cases in which $M$ is
approximately 0.024 in our code units (namely $\eta = 1.0507$ for
$\kappa =1/3$ and $\eta = 0.596693$ for $\eta = 0.08$).  For larger
values of $\kappa$, the horizon mass and angular momentum change less
after they are first formed, and settle down faster than for smaller
values of $\kappa$.}
\label{figure:horizons}
\end{figure}

For radiation fluids with $\kappa = 1/3$ we found that the black hole
masses increase by perhaps 10\% or so after they are first formed, and
that they settle down to equilibrium values rather quickly.  For
smaller values of $\kappa$, however, the black holes increase
significantly more, and it also takes them significantly longer
(measured in proper time at infinity), to settle down. In
Fig.~\ref{figure:horizons} we show examples for $\kappa = 1/3$ and
$\kappa = 0.08$, both for $\Omega = 0.3$ and both for cases in which
the horizon mass settles down to approximately 0.024 (in our code
units). This behavior makes it considerably more challenging to
analyze supercritical simulations for softer equations of
state. Fig.~\ref{figure:horizons} also shows that the black hole mass
and angular momentum appear to decrease slightly after having passed
through a maximum.  This is a numerical artifact that is related to
the finite numerical resolution of the black holes.  The effect is
more noticeable for smaller black holes, and contributes to the error
in determining the black hole parameters in particular close to the
black hole threshold.  In practice, we adopt the maximum values of the
black hole mass and angular momentum as our best approximations.

The scaling of the black hole angular momentum is mirrored by a
characteristic of subcritical evolutions, namely the maximum angular
velocity. In the axisymmetric spacetimes considered here, a
gauge-invariant measure of angular velocity is
\begin{equation} \label{omega}
\omega := \frac{\xi^a u_a}{\xi_a \xi^a} = \frac{u_\varphi}{g_{\varphi\varphi}},
\end{equation}
where $\xi = \partial/\partial \varphi$ is the rotational Killing
vector, and $u^a$ is the fluid four-velocity.  We evaluate $\omega$
for the fluid world-line at the center, $r = 0$.  For the initial data
at $t=0$, $\omega$ defined above agrees with the parameter $\Omega$
of the initial data.

\section{Numerical results}

From our two-parameter family of initial data with density parameter
$\eta$ and angular velocity parameter $\omega$, we {consider} both
``vertical" sequences for constant $\Omega$, {along which} we vary
$\eta$, and ``horizontal" sequences for constant $\eta$, {along which}
we vary $\Omega$.  These families appear as vertical and horizontal
lines in Figs.~\ref{figure:Omegaetaplane008} and
\ref{figure:Omegaetaplane1o3}, which illustrate the regions of
parameter space that we explore for $\kappa=0.08$ and $\kappa=1/3$,
respectively. Along each one of these sequences we locate the black
hole threshold, and record $M$ and $J$ for supercritical data, as well
as $\rho_{\rm max}$ and $\omega_{\rm max}$ for subcritical data. Most
of our horizontal sequences are at constant $\eta>\eta_{*0}$, but we
also take one horizontal sequence with constant $\eta<\eta_{*0}$,
which does not cross the black-hole threshold. The black hole
threshold itself is approximately given by the parabola
\begin{equation} 
\label{thresholdparabola}
C_0 \eta_* \simeq K (C_1 \Omega_*)^2
\end{equation}
(compare \cite{BaumgarteGundlach,gb16}) for the ranges of $\eta$ and
$\Omega$ that we have examined, even though small deviations can
be seen in Figs.~\ref{figure:Omegaetaplane008} and
\ref{figure:Omegaetaplane1o3} for large values of $\Omega$.

\subsection{Overview of different $\kappa$}

\subsubsection{Non-rotating data}

For non-rotating data with $\Omega = 0$ we have $Q = 0$ and hence,
from (\ref{deltaQP}), $\delta = 0$, independently of $\kappa$.  For
this special vertical sequence we may therefore adopt the power-law
scalings (\ref{powerlaws}) with 
\begin{equation}
P \simeq C_0(\eta-\eta_{*0})
\end{equation}
from Eqs.~(\ref{C0def}) and (\ref{Papproximation}) for any $\kappa$.
The power law (\ref{Mpowerlaw})
therefore becomes
\begin{equation}
M(\eta,0) \simeq \left(C_0 (\eta - \eta_{*0}) \right)^{\gamma_M}
\end{equation}
while (\ref{rhopowerlaw}) becomes
\begin{equation}
\rho_{\rm max}(\eta,0) \simeq c_\rho \left(C_0 (\eta_{*0} - \eta) \right)^{\gamma_\rho}.
\end{equation}
Fitting our numerical data to these scalings determines the
coefficients $C_0$, $c_\rho$ as well as the critical exponents
$\gamma_{M}$ and $\gamma_{\rho}$, which we tabulate in Table
\ref{Table:summary}.\footnote{{For these particular fits we fitted
    $\rho^{-1/2}_{\rm max}$ and $\omega^{-1}_{\rm max}$ to find
    exponents $- \gamma_\rho/2$ and $-\gamma_\omega$, which are listed
    in Table \ref{Table:summary}. The former in particular can then be
    compared directly with $\gamma_M$.}}  We find that our values
agree to within a few percent with both analytical \cite{Maison} and
previous numerical results \cite{NeilsenChoptuik}, and that
$\gamma_\rho = - 2 \gamma_M$ {within our estimated error}, as expected
from (\ref{crit_exponents}).

Several factors contribute to the error in our data.  In addition to
the finite-difference error in our numerical simulations, including
the uncertainties in determining the black hole mass and angular
momentum (see the discussion above), there is also some ambiguity in
what data to include in the fits to the power laws (\ref{powerlaws}).
Data too far away from the black hole threshold will no longer obey
the power laws, while data too close will be affected more strongly by
the finite-difference error for the very small structures formed
during the collapse.  This ambiguity alone leads to changes in the
critical exponents of about a few percent, which we therefore adopt as
an estimate of the error in our numerically determined critical
exponents.

\begin{table*}
\begin{tabular}{ll|llll|llll}
\hline
\hline
&		&	\multicolumn{4}{c|}{Subcritical Data}  & \multicolumn{4}{c}{Supercritical Data} \\ 
Fixed parameter &  Critical value	&  $- \gamma_\rho / 2$ & $- \gamma_\omega$	& $\lambda_0$  &$ \lambda_1$ & $\gamma_M$  & $\gamma_J$  & $\lambda_0$ & $\lambda_1$ \\
\hline
\hline
\multicolumn{2}{c|}{$\kappa = 0.5$}  	& 0.4774 	& 0.1061	& {\bf 2.095} 	& {\bf -0.778} 	& 0.4774 	& 1.326 	& {\bf 2.095} 	& {\bf -0.778}  	\\
\hline
$\Omega = 0$ & $\eta = 1.09588$ & 0.475 	& --  		& 2.11	& --  		& 0.484 	& -- 		& 2.07 	& -- \\
$\Omega = 0.01$ & $\eta = 1.0959$ 	& 0.476	& 0.103	& 2.05	& -0.789	& 0.497	& 1.31	& 2.01	& -0.636  \\
$\Omega = 0.1$ & $\eta = 1.0991$ 	& 0.488	& 0.103	& 2.05	& -0.789	& 0.486	& 1.28	& 2.01	& -0.634  \\
$\Omega = 0.3$ & $\eta = 1.1248$ 	& 0.482	& 0.109	& 2.07	& -0.774	& 0.497	& 1.30	& 1.97	& -0.616  \\
\hline
\hline
\multicolumn{2}{c|}{$\kappa = 1/3$}  	& 0.3558 	& 0.1779	& {\bf 2.811} 	& {\bf -0.5} 	& 0.3558 	& 0.8895 	& {\bf 2.811} 	& {\bf -0.5}  	\\
\hline
$\Omega = 0$ 	& $\eta = 1.01833$ 	& 0.356	& --		& 2.81	& --		& 0.358	& --		& 2.79	& -0.415  \\
$\Omega = 0.01$ & $\eta = 1.01836$ & 0.356	& 0.171	& 2.81	& -0.519	& 0.357	& 0.862	& 2.80	& -0.415  \\
$\Omega = 0.1$ & $\eta = 1.02198$ 	& 0.358	& 0.173	& 2.79	& -0.517	& 0.356	& 0.861	& 2.81	& -0.419  \\
$\Omega = 0.3$ & $\eta = 1.05058$ 	& 0.360	& 0.189	& 2.78	& -0.475	& 0.361	& 0.898	& 2.77	& -0.487  \\
\hline
\hline
\multicolumn{2}{c|}{$\kappa = 0.2$} 	& 0.2614	& 0.203	& {\bf 3.825}	& {\bf -0.222}	& 0.2614	& 0.581	& {\bf 3.825}	& {\bf -0.222} \\	
\hline
$\Omega = 0$	& $\eta = 	0.86926$	& 0.264	& --		& 3.79	& --		& 0.265	& --		& 3.78	& --	\\
$\Omega = 0.01$ & $\eta = 0.86930$& 0.265	& 0.197	& 3.78	& -0.257	& 0.264	& 0.561	& 3.79	& -0.125 \\	
$\Omega = 0.1$ & $\eta = 0.87381$	& 0.264	& 0.187	& 3.79	& -0.292	& 0.265	& 0.575	& 3.78	& -0.170 \\	
$\Omega = 0.3$ & $\eta = 0.90830$	& 0.269	& 0.185	& 3.72	& -0.312	& 0.271	& 0.628	& 3.69	& -0.317 \\	
\hline
\hline
\multicolumn{2}{c|}{$\kappa = 0.1$} 	& 0.1875	& 0.1932	& {\bf 5.333}	& {\bf 0.03} 	& 0.1875	& 0.3693	& {\bf 5.333} 	& {\bf 0.03} \\	
\hline
$\Omega = 0$ & $\eta = 0.62203$	& 0.192  	& --		& 5.21	& --		& 0.190 	& --		& 5.26	& --  \\
$\Omega = 0.01$ & $\eta = 0.62210$ & 0.193  	& 0.194	& 5.18	& 0.005	& 0.189 	& 0.360  	& 5.29	& 0.095 \\
$\eta = 0.6236$ & $\Omega = 0.049849$ & 0.197 & 0.178	& 5.08	&  -0.01	&  0.190	& 0.377	& 5.26	& 0.016 \\  
$\Omega = 0.05$ & $\eta = 0.62361$ & 0.196  	& 0.178	& 5.10	& -0.09	& 0.190	& 0.381	& 5.26	& -.005 \\
$\Omega = 0.1$ & $\eta = 0.62825$	& 0.197  	& 0.156	& 5.07	& -0.21	& 0.192	& 0.409	& 5.21	& -0.13 \\
$\Omega = 0.3$ & $\eta = 0.67234$	& 0.198  	& N/A	& 5.05	& N/A	& 0.192  	& 0.430	& 5.21	& -0.24 \\
\hline
\hline
\multicolumn{2}{c|}{$\kappa = 0.08$} 	& $ 0.174$	& $ 0.189$	& {\bf 5.75}	& {\bf 0.086} 	& $ 0.174$	& $ 0.333$	& {\bf 5.75} & {\bf 0.086} \\
\hline
$\Omega = 0$  & $\eta = 0.54302$		& 0.176	& --		& 5.68	& --		& 0.174	& --		& 5.75 	& --- \\
$\eta = 0.54309$ & $\Omega = 0.009809$ & 0.179	& 0.186	& 5.59	& 0.039	& 0.172	& 0.319 	& 5.81	& 0.145 \\
$\Omega = 0.01$ &	$\eta = 0.54309$ 	& 0.183	&  0.193	& 5.46	& 0.055	& 0.173	& 0.317	&  5.78 	&  0.167 \\
$\eta = 0.5447$ & $\Omega = 0.04922$	& 0.191	& 0.166	& 5.24	& -0.13	&0.174	& 0.337	& 5.75	& 0.063 \\
$\Omega = 0.05$ &	$\eta = 0.54475$ 	& 0.183	& 0.162 	& 5.46	& -0.11	& 0.174	& 0.341	& 5.78 	& 0.040\\
$\Omega = 0.1$ &	$\eta = 0.54982$ 	& 0.178	& 0.169 	& 5.61	& -0.05	& 0.178	& 0.353	& 5.61 	& 0.017\\
$\Omega = 0.3$ &	$\eta = 0.59669$ 	& 0.189	& N/A 	& 5.29	& N/A	& 0.178	& 0.389	& 5.61 	& -0.18\\
\end{tabular} 
\caption{ For each value of $\kappa$ we list, in bold face, the
  analytical values of $\lambda_0$ (from \cite{Maison}) and
  $\lambda_1$ (Eq.~(\ref{lambda1theory})) in the same row. For $\kappa
  = 0.08$, no analytical value for $\lambda_0$ is available, but, from
  interpolation, we approximate this value to be about 5.75.  Also
  listed in these rows are the critical exponents $\gamma$ as computed
  from (\ref{crit_exponents}) obtained from fits to the expressions
  (\ref{scaling}). All other rows contain our numerical data for the
  critical exponents $\gamma$ for the corresponding values of
  $\kappa$.  From these, we compute heuristic values of $\lambda_0$
  and $\lambda_1$ from Eqs.~(\ref{lambdasubcritical}) for subcritical
  data or (\ref{lambdasupercritical}) for supercritical data.  Entries
  N/A indicate that the fits were too poor to provide an accurate
  estimate for the critical exponents {(see, e.g., Fig.~\ref{figure:omegascaling} below)}.  We list three significant
  digits for all numerically determined exponents, even though we
  estimate their relative errors to be on the order of a few percent.
}
\label{Table:summary}
\end{table*}

We next analyze both vertical and horizontal sequences for rotating data.  

\subsubsection{Rotating data}

For $\kappa > 1/9$, we expect that, in the vicinity of the black hole
threshold, all characteristic variables can be described by the
power-law scalings (\ref{powerlaws}).  We approximate 
\begin{equation}
\label{tildePvert}
\tilde P \simeq C_0 ( \eta - \eta_*)
\end{equation}
on our vertical sequences,
\begin{equation}
\label{tildePhor}
\tilde P \simeq K C_1^2 ( \Omega^2 - \Omega_*^2) {\simeq 2 K C_1^2
  \Omega_* ( \Omega - \Omega_*)}
\end{equation}
on horizontal sequences with $\eta>\eta_{*0}$, 
\begin{equation}
\tilde P\simeq C_0(\eta-\eta_{*0})-K(C_1\Omega)^2
\end{equation}
on horizontal sequences with $\eta<\eta_{*0}$, and on all of these
\begin{equation}
\label{Qboth}
Q\simeq C_1\Omega.
\end{equation}
Inserting these expressions into (\ref{powerlaws}) we can make fits to
our data to obtain the critical exponents (\ref{crit_exponents})
together with the threshold parameters $\eta = \eta_*(\Omega)$ and
$\Omega = \Omega_*(\eta)$. We list our results in Table
\ref{Table:summary}. As expected, the threshold data lie approximately
on the parabola (\ref{thresholdparabola}).

From eqs.~(\ref{crit_exponents}) we see that we can also compute
``heuristic" values of $\lambda_0$ and $\lambda_1$ from the
numerically determined critical exponents.  For supercritical data we
find
\begin{subequations} \label{lambdasupercritical}
\begin{eqnarray} 
\lambda_0 & = & 1 / \gamma_M \\
\label{lambda1supercritical}
\lambda_1 & = & 2 - \gamma_J/\gamma_M.
\end{eqnarray}
\end{subequations}
while for subcritical data we find 
\begin{subequations} \label{lambdasubcritical}
\begin{eqnarray} 
\lambda_0 & = & - 2 / \gamma_\rho \\
\label{lambda1subcritical}
\lambda_1 & = & 2 \gamma_\omega/\gamma_\rho - 1,
\end{eqnarray}
\end{subequations}
The resulting heuristic values for $\lambda_0$ and $\lambda_1$ are
also included in Table \ref{Table:summary}.

We have found qualitatively similar effects for all $\kappa > 1/9$,
but, for concreteness, will focus on $\kappa = 0.2$ here.  

\begin{figure}
\ifprintfigures
\includegraphics[width=3.7in]{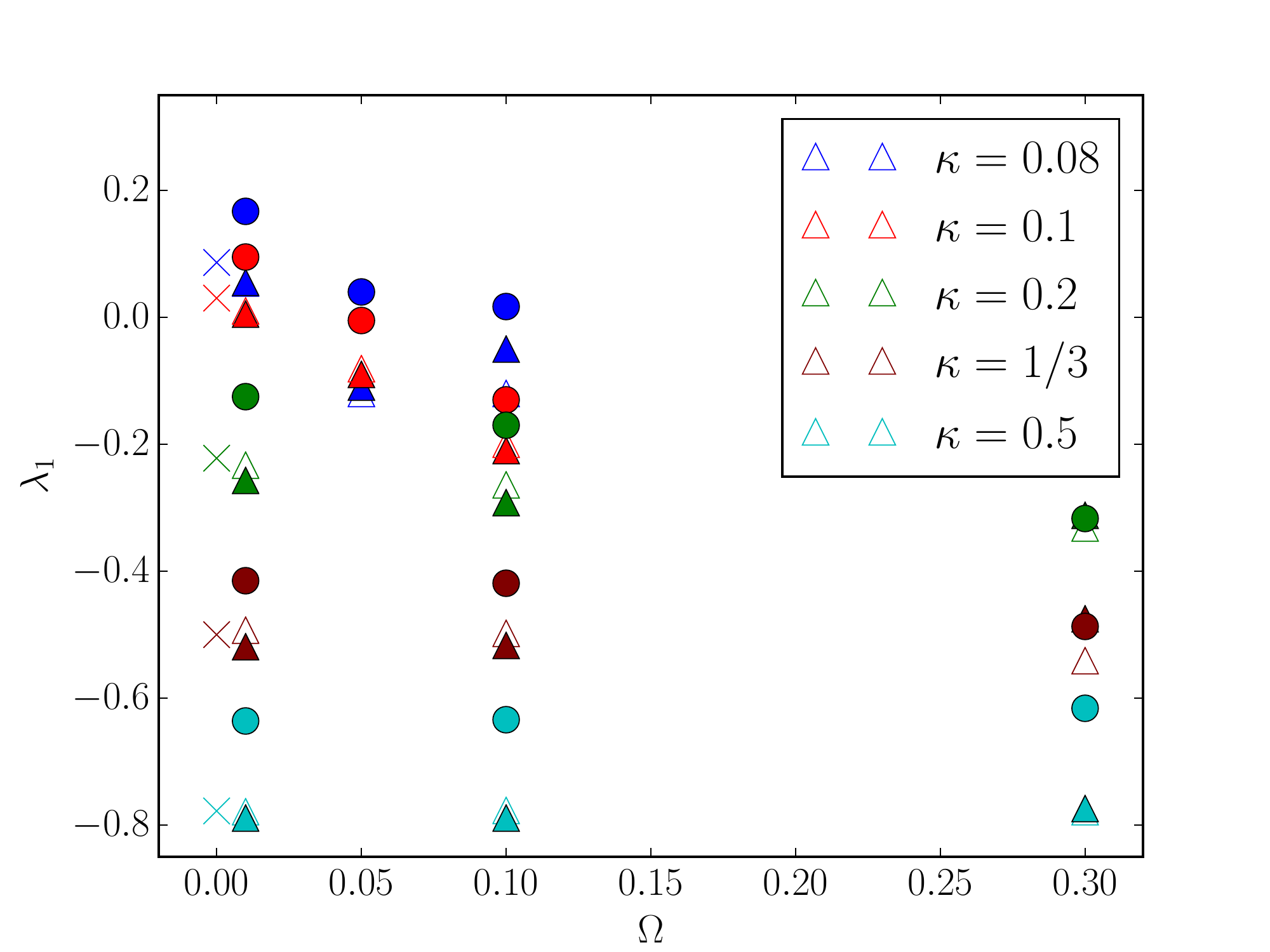}
\fi
\caption{Theoretical and heuristic values of $\lambda_1$.  The
  crosses, plotted at $\Omega = 0$, represent the theoretical values
  given by (\ref{lambda1theory}).  Filled circles (triangles)
  represent the results from power-law fits for the supercritical
  (subcritical) data tabulated in Table \ref{Table:summary}.  Open
  triangles represent dynamically determined values, obtained from
  fits to (\ref{rho_self_sim}) and (\ref{omega_self_sim}) for
  near-critical evolutions.  The subcritical and dynamical heuristic
  values agree well with the theoretical values for small $\Omega$
  (and often cannot be distinguished in the figure).  The
  supercritical heuristic values appear to overestimate $\lambda_1$,
  but, as discussed in the text, are also affected by a larger error.
  The heuristic values for $\lambda_1$ appear to be smaller than the
  theoretical ones for larger $\Omega$, in particular for soft
  equations of state with small $\kappa$.  As discussed in the text,
  several sources of error contribute to the uncertainty in these
  numbers, which therefore should be taken as estimates only.}
 \label{figure:lambda_1}
\end{figure}

For small $\Omega$, we find good agreement of all exponents $\gamma$
with the respective analytical values.  Conversely, the heuristic
values for $\lambda_0$ and $\lambda_1$ agree {well} with the
theoretical ones. We note that the results for $\lambda_1$ have a
larger {\em relative} error, when compared with the analytical values,
than the exponents $\gamma$.  This can be understood from the fact
that $\lambda_1$ is relatively small, but, in
  (\ref{lambda1supercritical}) and (\ref{lambda1subcritical}), is
computed from the difference of two numbers of order unity.  Assuming
that each $\gamma$ has a relative error of 5\%, say, will result in an
{\em absolute} error in $\lambda_1$ of about 0.15 when computed from
supercritical data, and about 0.05 when computed from subcritical
data.  Our data are within this range, at least for small $\Omega$.
Consistent with these estimates we find that, for small $\Omega$, our
heuristic values for $\lambda_1$ agree quite well with the theoretical
values when computed from the subcritical data, but are somewhat
larger than the theoretical values when computed from the
supercritical data.  We also show these results in
Fig.~\ref{figure:lambda_1}, where supercritical (subcritical) data are
shown as filled circles (triangles), and the theoretical values as
crosses.

Interestingly, however, we find that the fitted critical exponents
change slightly when we increase $\Omega$.  For $\gamma_\rho$ and
$\gamma_M$ these changes are quite small, and it is not clear whether
they can be distinguished from numerical error.  For $\gamma_{\omega}$
and $\gamma_J$, however, the changes are somewhat larger, and they
appear to increase, for a given $\Omega$, with decreasing $\kappa$.
For a radiation fluid, with $\kappa = 1/3$, we also found a small
increase in $\gamma_J$ with $\Omega$ (see Table I in
\cite{BaumgarteGundlach}), but all changes were well within the
estimate of the numerical error.  For $\kappa = 0.2$, the changes
become more noticeable, with both $\gamma_\omega$ and $\gamma_J$
increasing for larger $\Omega$.  Correspondingly, we find that the
heuristic values of $\lambda_1$ also appear to change with $\Omega$.
For both subcritical and supercritical data we found that the
heuristic values of $\lambda_1$ decrease with increasing $\Omega$,
i.e.~become more negative.   {This trend can also be seen in Fig.~\ref{figure:lambda_1}.}

We note that \cite{ChoHLP03} also reported
that, in non-spherical deformations of critical collapse with a scalar
field, the critical exponent as well as the period of the discretely
self-similar critical solution depend on the size of the deformation
(see their Table I).

\begin{figure}
\ifprintfigures
\includegraphics[width=3.75in]{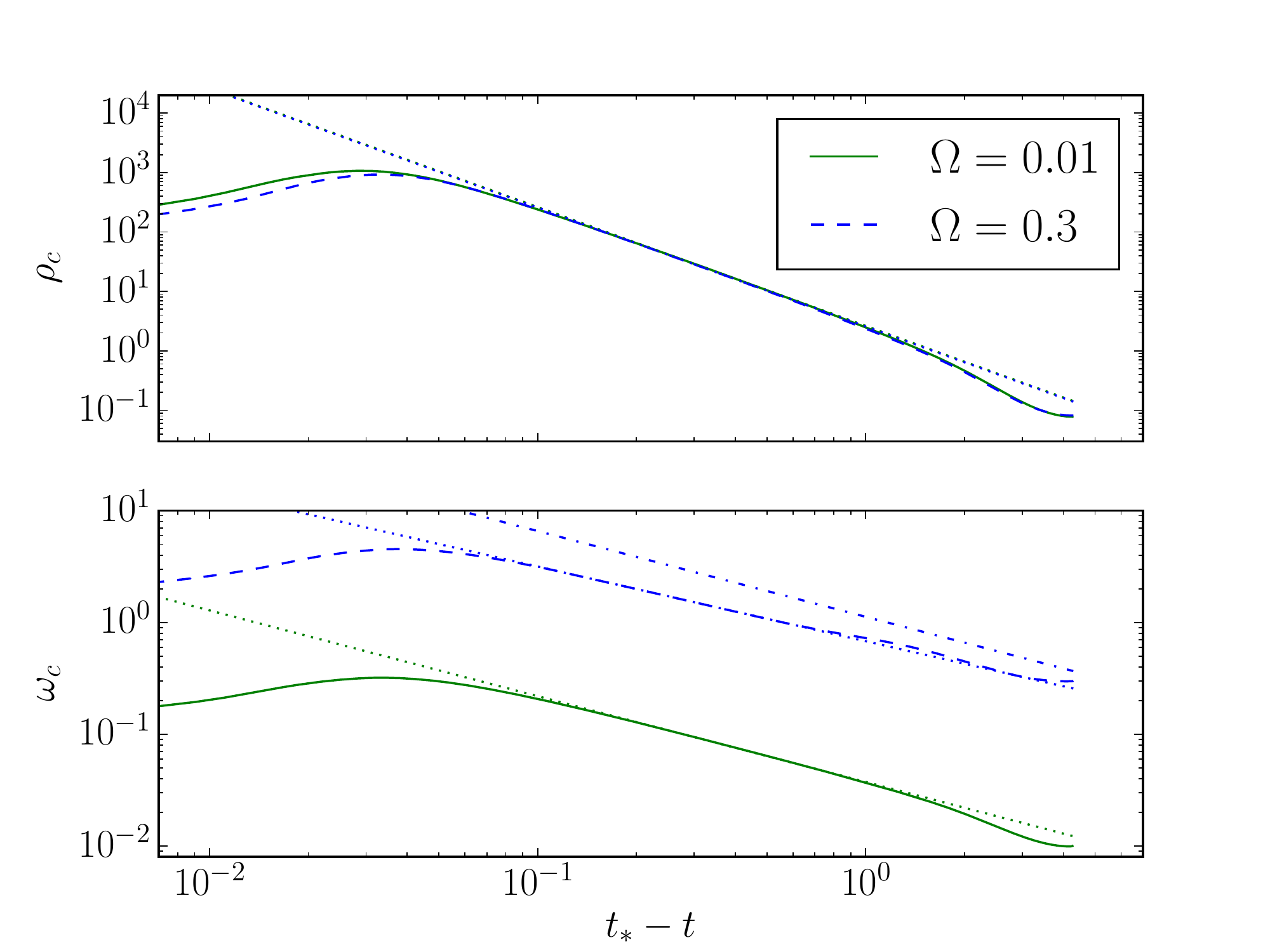}
\fi
\caption{The central density $\rho_c$ (upper panel) and central
  rotation rate $\omega_c$ (lower panel) for subcritical evolution
  close to the black hole threshold as a function of central proper time $t$, 
  {for $\kappa = 0.2$}.
  Time advances to the left, as $t$ approaches the time $t_*$ of the
  accumulation event. The solid and dashed lines are numerical data,
  while the dotted lines are the fits (\ref{rho_self_sim}) and
  (\ref{omega_self_sim}) for the evolution during Phase~2.
  For comparison, we show the fit for $\Omega=0.01$ scaled up by
  a factor of $30$ in $\Omega$ to $\Omega=0.3$, using the same
  heuristic $\lambda_1$ and $d_\omega$  (the dash-dotted line). 
  This shows that over the
  dynamical range of $t_*-t$ that we see here, the increase in
  $\Omega$ appears to affect an overall scale in $\omega$ more than
  its scaling power.}
\label{figure:self_sim_fit}
\end{figure}

To our surprise we found that, at least for modest values of $\Omega$,
the above scaling laws model data quite accurately even for $\kappa <
1/9$.  In the presence of two unstable modes, when $\lambda_1 > 0$ and
hence $\epsilon > 0$, we can no longer assume that $\delta \to 0$ at
the black hole threshold (see Eq.~(\ref{deltaQP})).  Accordingly,
we
are now probing the scaling functions for potentially large $\delta$,
so that the leading-order approximations
(\ref{Fleading}) for the
scaling functions may no longer apply, meaning that we also can no
longer assume the power-law scalings (\ref{powerlaws}) to hold.
Nevertheless, we found surprisingly good agreement, {especially 
if heuristic values are adopted for $\lambda_1$ which, as for $\kappa > 1/9$, 
decrease with increasing $\Omega$.}

\subsubsection{Determination of $\lambda_1$ from time evolution}

As an alternative approach to estimating an ``effective" value of
$\lambda_1$ we also examined the time evolution of the central density
$\rho_c$ and central angular velocity $\omega_c$ for near-critical
initial data. Two examples, for $\kappa = 0.2$ with $\Omega = 0.01$
and $\Omega = 0.3$, are shown in Fig.~\ref{figure:self_sim_fit}.  In
the upper panel we show the evolution of the central density.  We plot
the data as a function of $t_* - t$, where $t_*$ is the proper time of
the accumulation event, so that time advances from right to left in
the figure.  The three phases of the evolution, which we described
earlier, are clearly visible.  In Phase 1 the data approach the
critical solution.  During Phase 2, the evolution follows the critical
solution.  On dimensional grounds, the density must be
  approximated by
\begin{equation} \label{rho_self_sim}
\rho_c(t) \simeq d_\rho \, (t_* - t)^{-2}
\end{equation}
with $d_\rho$ a universal constant during this part of the
evolution. Fits to this scaling are included as the dotted line
  in the upper panel of Fig.~\ref{figure:self_sim_fit}.  Phase 2 ends
when perturbations of the critical solution become large; for the
subcritical evolutions shown in Fig.~\ref{figure:self_sim_fit} the
density drops below that of the critical solution.  This is clearly
visible in the left part of the figure.

The central rotation rate $\omega_c$, shown in the lower panel of
Fig.~\ref{figure:self_sim_fit} would, from dimensional analysis
  alone, be expected to scale as $(t_* - t)^{-1}$.  In addition,
however, it also grows or decays with the mode $Z_1$ as
$\exp\lambda_1 \tau \propto (t_* - t)^{\lambda_1}$. It should
therefore be approximated by
\begin{equation} \label{omega_self_sim}
\omega_c \simeq Q d_\omega \, (t_* - t)^{-1 + \lambda_1}.
\end{equation}
with $d_\omega$ a universal constant. In practice, we can now fit
$\rho_c$ during Phase 2 to the fit (\ref{rho_self_sim}); this
determines $d_\rho$ and $t_*$.  We then adopt this value of $t_*$ in a
fit of $\omega_c$ to (\ref{omega_self_sim}). This allows us to
determine $\lambda_1$ from the dynamical data. Admittedly these 
results for $\lambda_1$ have a significant error. One significant
source of error is the time window over which the fits are performed,
which should be restricted, of course, to Phase 2. We can nevertheless
obtain estimates for $\lambda_1$, and interestingly these estimates
are consistent with our observations based on the critical exponents
described above.

The examples shown in Fig.~\ref{figure:self_sim_fit} are for $\kappa =
0.2$ with $(\eta,\Omega) = (0.86930346, 0.01)$, and $(
0.9082978,0.3)$.  The fit (\ref{rho_self_sim}) for $\rho$ captures
properties of the unique critical solution, and therefore takes the
same shape for both evolutions shown in the figure.  For $\omega$ we
can also identify an evolution well approximated by the fit
(\ref{omega_self_sim}) during Phase 2; however, even in the figure we
notice that the two curves have a slightly different slope.  For
$\Omega = 0.01$ our fits suggest $\lambda_1 = -0.23$, while for
$\Omega = 0.3$, we obtain $\lambda_1 = -0.33$. These values are quite
similar to those obtained form the power-law fits to $\gamma_\rho$ and
$\gamma_\omega$ discussed above, and listed in Table
\ref{Table:summary}.

Using similar fits we estimated effective {dynamical} values of
$\lambda_1$ for different values of $\kappa$ and $\Omega$; they are
shown as {open} triangles in
Fig.~\ref{figure:lambda_1}.  In general, the trends discussed above
appear to hold.  {The dynamical estimates for $\lambda_1$ agree
  quite well with those obtained from subcritical fits (filled
  triangles in Fig.~\ref{figure:lambda_1}).}  For small $\Omega$, we
{also} find good agreement with the {theoretical} values
(\ref{lambda1theory}), shown as crosses in Fig.~\ref{figure:lambda_1}.
For larger values of $\Omega$ however, the effective values appear to
be smaller than the {theoretical} values.  The changes are
larger for smaller values of $\kappa$; in particular, for $\kappa =
1/3$ and 0.5 any changes appear to be well within the error of the
data.

\subsection{Example $\kappa=0.08$}

As a specific example of rotating collapse with $\kappa<1/9$, we focus
on $\kappa = 0.08$ in the remainder of this Section.  Even
though we do not have an analytical value for $\lambda_0$ for this
value of $\kappa$, it seems to be a good compromise between two
competing effects: for smaller $\kappa$, the numerical simulations
become increasingly challenging (see Section \ref{sec:diagnostics}
above), and for larger $\kappa$, closer to 1/9, $\epsilon$ becomes
smaller.

\begin{table}
\begin{tabular}{l|l}
parameter & status \\
\hline
\hline
{$\lambda_0=1/\gamma_M \simeq 5.78$} &  \\
$c_\rho\simeq 0.0065$ & universal but fitted\\ 
$c_\omega\simeq 0.144$ & \\
{$\epsilon = \lambda_1 / \lambda_0 \simeq 0.015$ }\\
\hline
$\lambda_1=7/81\simeq 0.0864198$ & exact value used \\
\hline
$C_0\simeq 0.0004$ &  \\ 
$C_1\simeq 2.15$ & family-dependent \\
$K\simeq 0.00006$ & \\  
\hline
\end{tabular}
\caption{The parameters used in computing the theoretical
fits {for $\kappa = 0.08$} in the following figures.}
\label{table:fittingparameters}
\end{table}

\begin{table}
\begin{tabular}{l|l|l}
fixed parameter & critical parameter & color \\
\hline
\hline
$\eta=0.543$ & all subcritical & orange \\
\hline
$\Omega=0$ & $\eta_*\simeq0.54302320$ & black \\
\hline
$\Omega=0.01$ & $\eta_*\simeq0.5430926$ & grey \\
$\eta=0.54309$ & $\Omega_*\simeq0.009809$ & purple \\
\hline
$\Omega=0.05$ & $\eta_*\simeq 0.54475323$ & cyan \\
$\eta=0.5447$ & $\Omega_*\simeq0.049219789$ & red \\
\hline
$\Omega=0.1$ & $\eta_*\simeq0.54981506$ & blue \\
$\Omega=0.3$ & $\eta_*\simeq0.59669136$ & green \\
\hline
\end{tabular}
\caption{Overview of the one-parameter sequences of data evolved for
  $\kappa=0.08$. The grey and purple, and the cyan and red, families
  respectively cross the black hole threshold at almost the same
  points. The critical values are used to evaluate $\ln|\tilde
    P|$ on the horizontal axes of the following figures. We will plot
  supercritical data as filled circles, and subcritical data as filled
  triangles.}
\label{table:sequences}
\end{table}

We start by determining the different family-dependent parameters from
the different power laws.  Specifically, we initially determine
$\eta_{*0}$ and $C_0$ from $M(\eta)$ at $\Omega=0$, $C_1$ from
$J(\eta,\Omega)$ at small $\Omega$, and $K$ from the shape of the
black hole threshold. We then fit to the expressions
(\ref{powerlaws}).  From these fits we determine the critical
exponents as well as the threshold parameters $\eta_*$ and
$\Omega_*$.  The estimated values of the
universal parameters for $\kappa=0.08$, and of the family-dependent
parameters of our {two}-parameter family of initial data are given in
Table~\ref{table:fittingparameters}. Our {one}-parameter sequences of
data, together with their critical parameter values and color-codings
are given in Table~\ref{table:sequences}. In all of the following
plots we will denote supercritical data with filled dots and
subcritical data with filled triangles.

\subsubsection{Limits on $\alpha_*$}

As mentioned above, we have hidden our ignorance of $\alpha_*$ [or
  $(\delta_*,s_*)$] by using a degeneracy between the overall scale
$\tau$ and the scaling functions $F$. This means that we will not be
able to obtain information about $\alpha_*$ directly from the scaling
laws. However, we do get a lower bound on $\delta_*$ from the fact
that the black-hole threshold is approximately a parabola in the
entire region of the $\Omega\eta$-plane that we have surveyed, as
follows.  In Fig.~\ref{figure:Omegaetaplane008} we have
plotted contours of $\delta$, assuming that the leading{-order}
approximations (\ref{PQapproximations}) are exact, and using our
numerically determined values of $C_0$, $C_1$ and $K$. We see that for
any value of $|\delta_*|<0.8$ the black hole threshold would look
qualitatively different, with the parabola either turning up at a
sharp corner into an essentially vertical line (for $0<\delta_*<0.8$),
or even turning sharply down into an essentially vertical line (for
$-0.8<\delta_*<0$). We do not {observe} either {behavior} in the
region of parameter space we have explored, and so we know that
$\delta_*>0.8$, with either sign of $s_*$ possible.  (This lower limit
on $|\delta_*|$ corresponds to an upper limit on $|\alpha_*-(\pi/2)|$,
but that limit also depends on the unknown value of $b_1$.)

\subsubsection{Leading-order power laws}

\begin{figure}
\ifprintfigures
\includegraphics[width=3.3in]{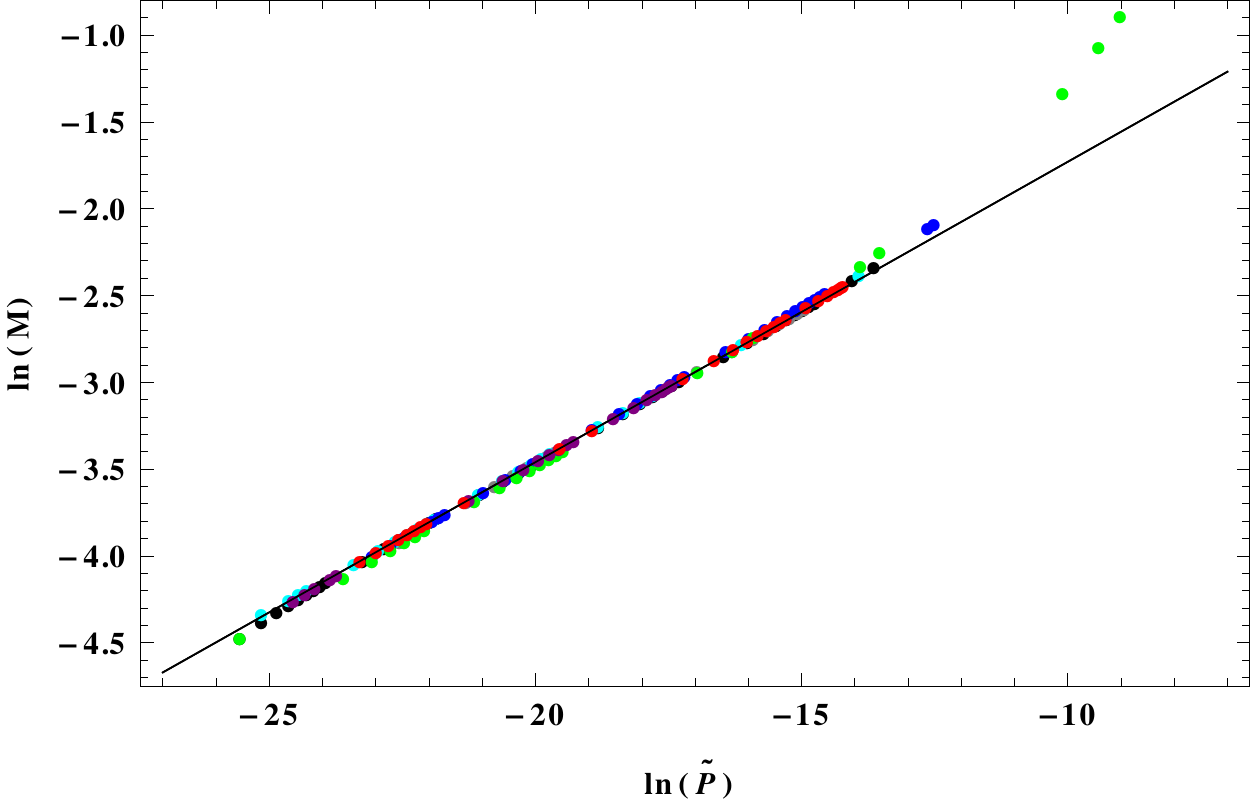}
\fi
\caption{Leading-order scaling of the black-hole mass $M$. The colored
  dots show numerical results for $\ln M$, plotted against $\ln\tilde
  P $. $\tilde P(\Omega,\eta)$ has been computed from
  (\ref{tildePvert}) for vertical sequences and from (\ref{tildePhor})
  for horizontal sequences. The parameters and color codings are given
  in Tables~\ref{table:fittingparameters} and
  \ref{table:sequences}. The solid black line is the theoretical
  prediction.}
\label{figure:Mscaling}
\end{figure}

We test the expected leading-order scalings (\ref{scaling}) in
Figs.~\ref{figure:Mscaling} {through} \ref{figure:omegascaling}.  We
plot data and predictions for $M$, $J/\Omega$, $\rho_{\rm max}$ and
$\omega_{\rm max}/\Omega$ against $\ln|\tilde P|$. 

\begin{figure}
\ifprintfigures
\includegraphics[width=3.3in]{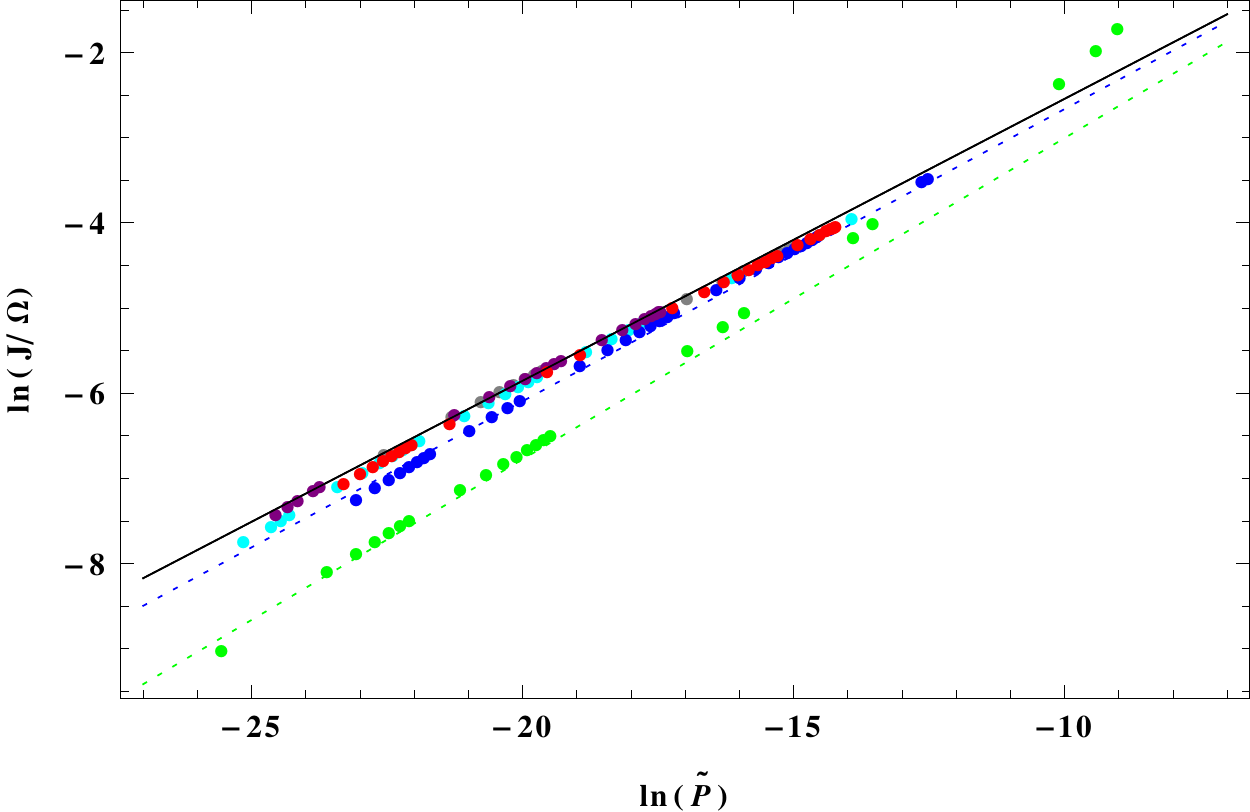}
\fi
\caption{Leading-order power-law scaling of the black hole angular
  momentum divided by the initial angular velocity, $J/\Omega$.  The
  solid line black line shows the theoretical prediction given by the
  theoretical value $\lambda_1 = 7/81$ (see
  Table~\ref{table:fittingparameters}), while the dotted colored lines
  show theoretical predictions with ad-hoc values of
  $\lambda_1=0.017$ and $-0.18$ for the vertical sequences
  $\Omega=0.1$ (blue) and $\Omega=0.3$ (green).}
\label{figure:Jscaling}
\end{figure}

\begin{figure}
\ifprintfigures
\includegraphics[width=3.3in]{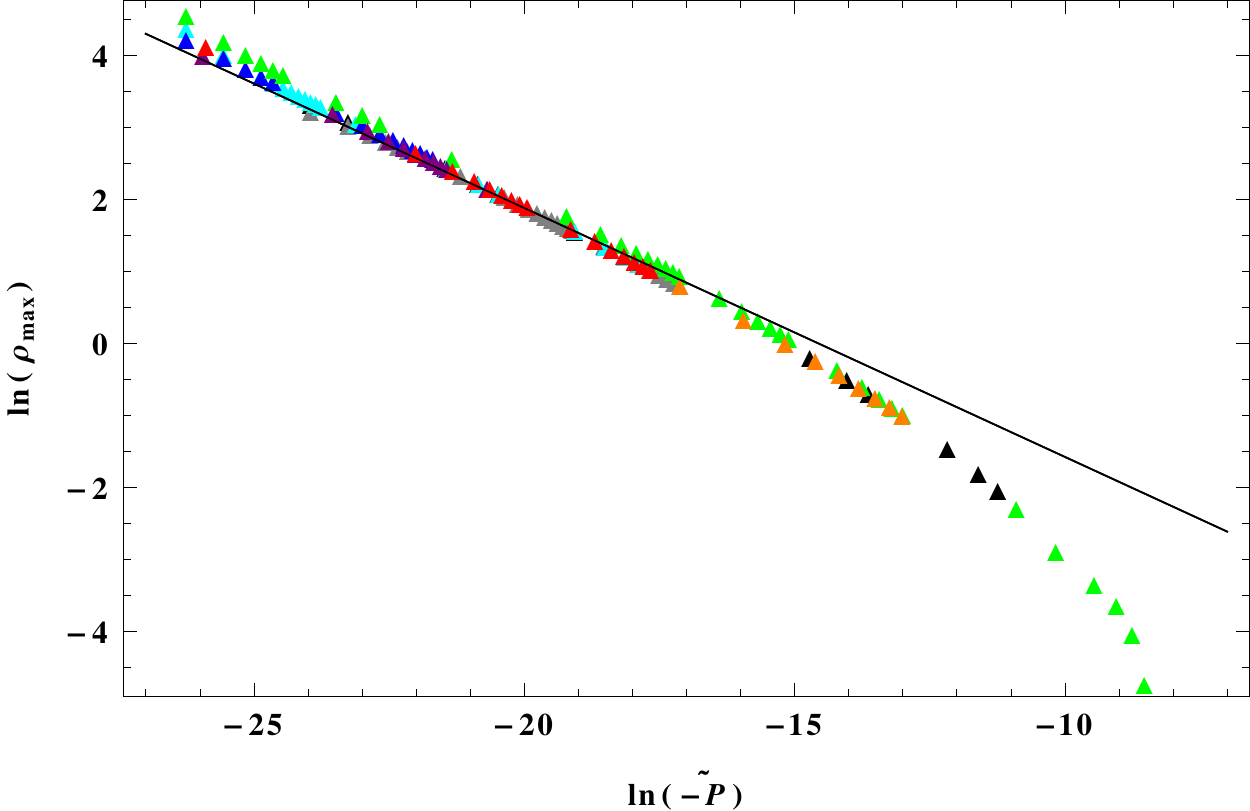}
\fi
\caption{Leading-order power-law scaling of the maximum central
  density $\rho_{\rm max}$, now for the subcritical halves of the same
  sequences of initial data, and hence plotted against $\ln(-\tilde P)$.}
\label{figure:rhoscaling}
\end{figure}

\begin{figure}
\ifprintfigures
\includegraphics[width=3.3in]{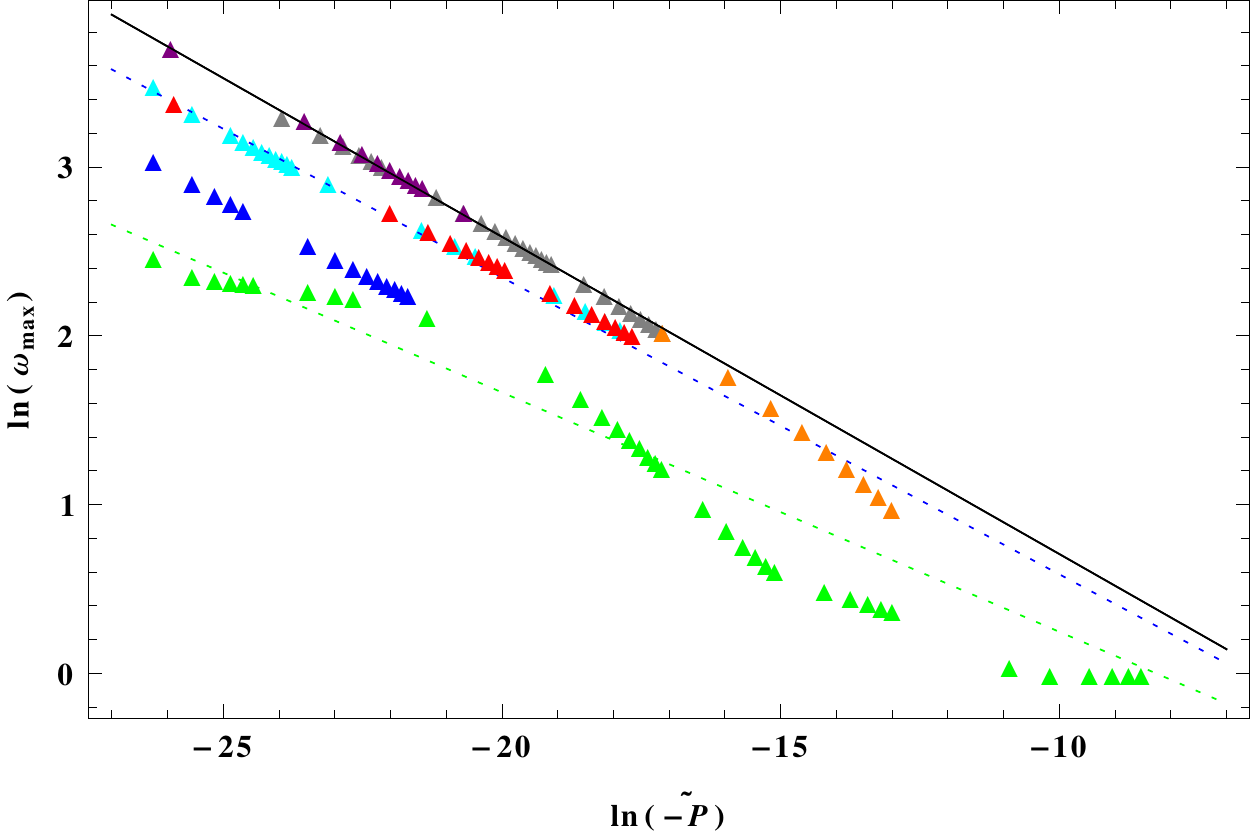}
\fi
\caption{Leading-order power-law scaling of the maximum central
  angular velocity, divived by the initial angular velocity,
  $\omega_{\rm max}/\Omega$. The dotted coloured lines use the same
  heuristic $\lambda_1$ as in Fig.~\ref{figure:Jscaling}.}
\label{figure:omegascaling}
\end{figure}

The first thing that stands out is how good the leading{-order}
power-law fits (with the exact $\lambda_1$ and the leading-order
approximations to the scaling functions) are for most of our data.  We
do see some clear deviations from the power laws far from the black
hole threshold (towards the right), where we have numerical data only
for the $\Omega=0$ (black) and $\Omega=0.3$ (green) sequences. We are
content to dismiss these as the theory breaking down far from the
black hole threshold.

We do not see any clear systematic deviation close to the black hole
threshold (towards the left) that would indicate nontrivial universal
scaling functions for $M$ or $\rho$.  We do, however, see that the
numerical data for both $J$ and $\omega_{\rm max}$ at large initial
angular momentum, namely for $\Omega=0.1$ (blue) and $\Omega=0.3$
(green), are both lower and steeper than the theoretical predictions.
As discussed above, in the context of Table~\ref{Table:summary} for
different equation of state parameters $\kappa$, both effects can be
captured by assuming, in a purely heuristic manner, that an effective
value of $\lambda_1$ decreases with {increasing} $\Omega$.  We show
this in Figs.~\ref{figure:Jscaling} and \ref{figure:omegascaling} in
the dashed theoretical lines.  For the supercritical data we find good
agreement assuming heuristic values of $\lambda_1 = 0.017$ for $\Omega
= 0.1$ and $\lambda_1 = -0.18$ for $\Omega = 0.3$; the same values as
those listed in Table \ref{Table:summary}.  (Note that we have
  not adjusted the value of $C_1$.)
 
Finally, the data for $\Omega=0.3$ (green) do no longer appear to
follow a power law.  The departure from a power law is very clearly
visible for $\omega_{\rm max}$, but is also present in $M$, $J$ and
$\rho_{\rm max}$.  We have no plausible theoretical interpretation for
this behavior.  It is possible that these data are sufficiently far
away from the critical solution that predictions based on linear
perturbations of the critical solutions fail, but we also cannot rule
out that this behavior represents a numerical artifact.  Accordingly,
we did not list values for $\gamma_{\omega}$ in
Table~\ref{Table:summary} for $\Omega = 0.3$ and $\kappa < 1/9$.

\subsubsection{Breakdown of scaling?}

\begin{figure}
\ifprintfigures
\includegraphics[width=3.3in]{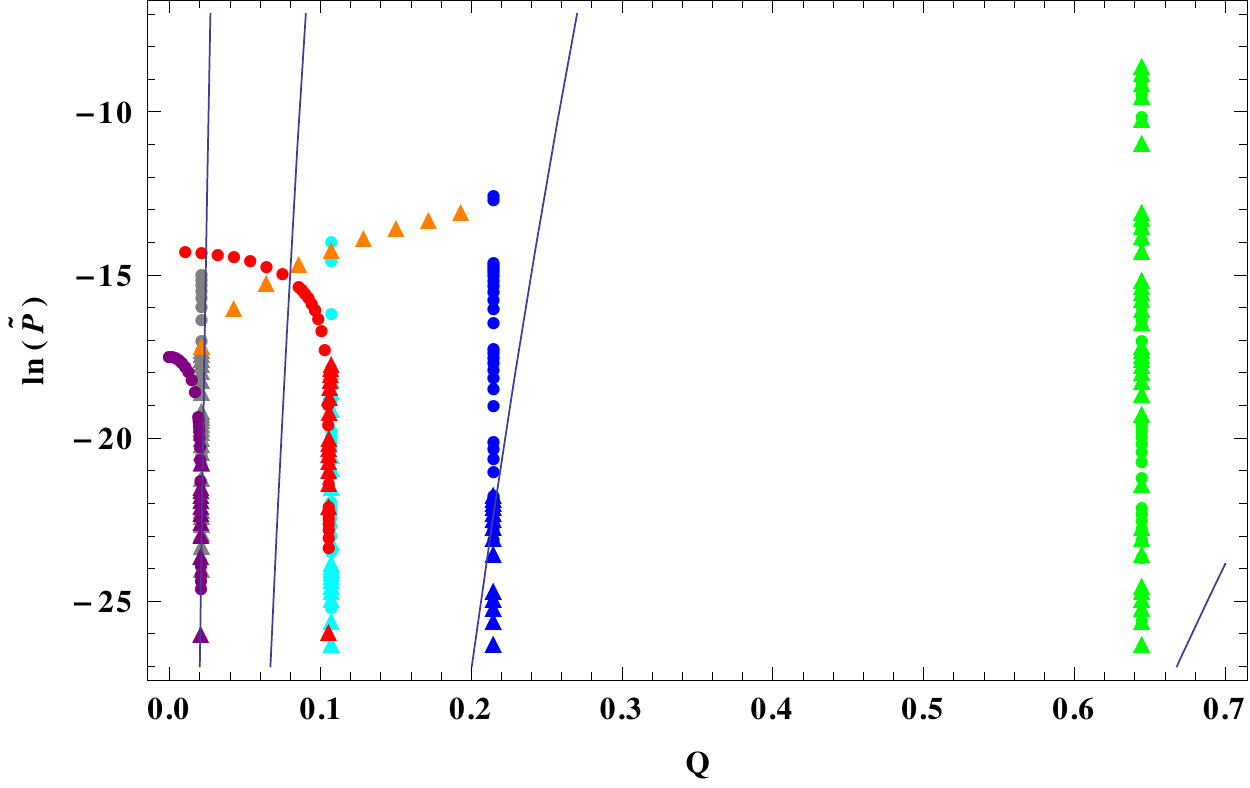}
\fi
\caption{Our initial data in the $Q\tilde P$-plane. As $|\tilde P|$
  becomes very small near the black-hole threshold, we plot
  $\ln|\tilde P|$ rather than $\tilde P$. Different one-parameter
  sequences of initial data are colour-coded as in
  Table~\ref{table:sequences}, with dots denoting supercritical data
  and triangles subcritical data. The solid curves are given by
  $|\tilde P|=|Q/b_1|^{1/\epsilon}$, for $b_1=0.03$, $0.1$, $0.3$ and
  $1$ (from left to right). Hence all our data obey $|\tilde
  P|>|Q/b_1|^{1/\epsilon}$ for $b_1\ge 1$.}
\label{figure:QPplot}
\end{figure}

{We do not see any breakdown of scaling. To get a least some
  idea whether we should expect one} at our level of fine-tuning, we
replace the condition (\ref{scalingbreakdown}) for scaling to break
down by $|\tilde P|<|Q/b_1|^{1/\epsilon}$. We do not know the value of
$b_1$, but Fig.~\ref{figure:QPplot} shows that this latter condition
is not met for any of our initial data as long as $b_1>1$. (We have
used the exact value of $\lambda_1$ in $\epsilon$ here. {For a
  smaller but still positive heuristic $\lambda_1$ even more
  fine-tuning would be required, while for $\lambda_1<0$ we do not
  expect scaling to break down at all.)}

\subsubsection{Beyond leading order}

As we discussed above, our numerical data can be fit to power laws
(\ref{powerlaws}) quite well, at least for modest values of $\Omega$,
but these fits require adopting heuristic and $\Omega$-dependent
values of the Lyaponov exponent $\lambda_1$.  We now ask
whether, alternatively, the observed behavior of the numerical data
can be explained by higher-order terms in the scaling functions
$\tilde F_J(\tilde\delta)$ and $\tilde F_\rho(\tilde\delta)$.  In
the following we will therefore adopt the theoretical value
$\lambda_1 = 7/81$ (see Eq.~(\ref{lambda1theory})) and will examine
deviations of the numerical data from power laws.

\begin{figure}
\ifprintfigures
\includegraphics[width=3.3in]{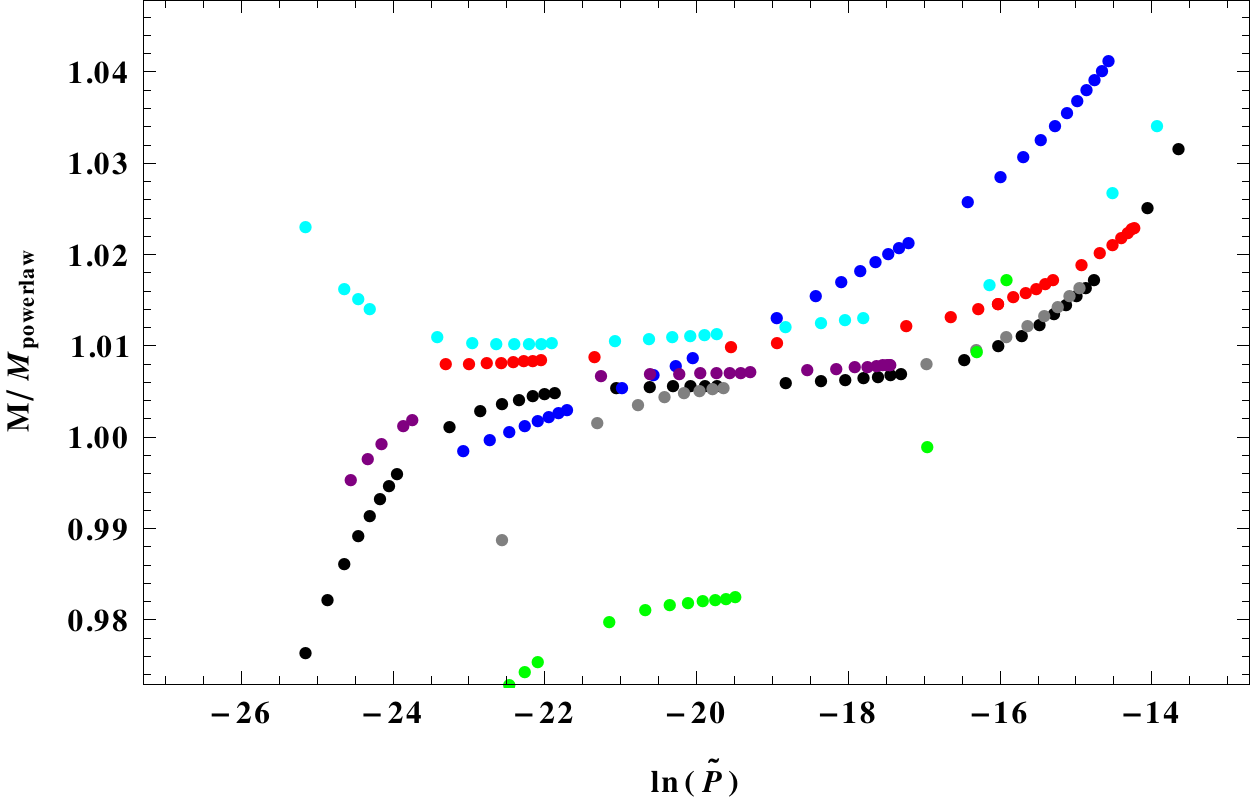}
\fi
\caption{{Black hole masses} $M$ with the expected leading-order power
  law taken out, plotted against $\ln \tilde P$. We have cut off data
  far from the black hole threshold, where the deviations from the
  expected power law are large (compare Fig.~\ref{figure:Mscaling}).}
\label{figure:Mresidual}
\end{figure}

\begin{figure}
\ifprintfigures
\includegraphics[width=3.3in]{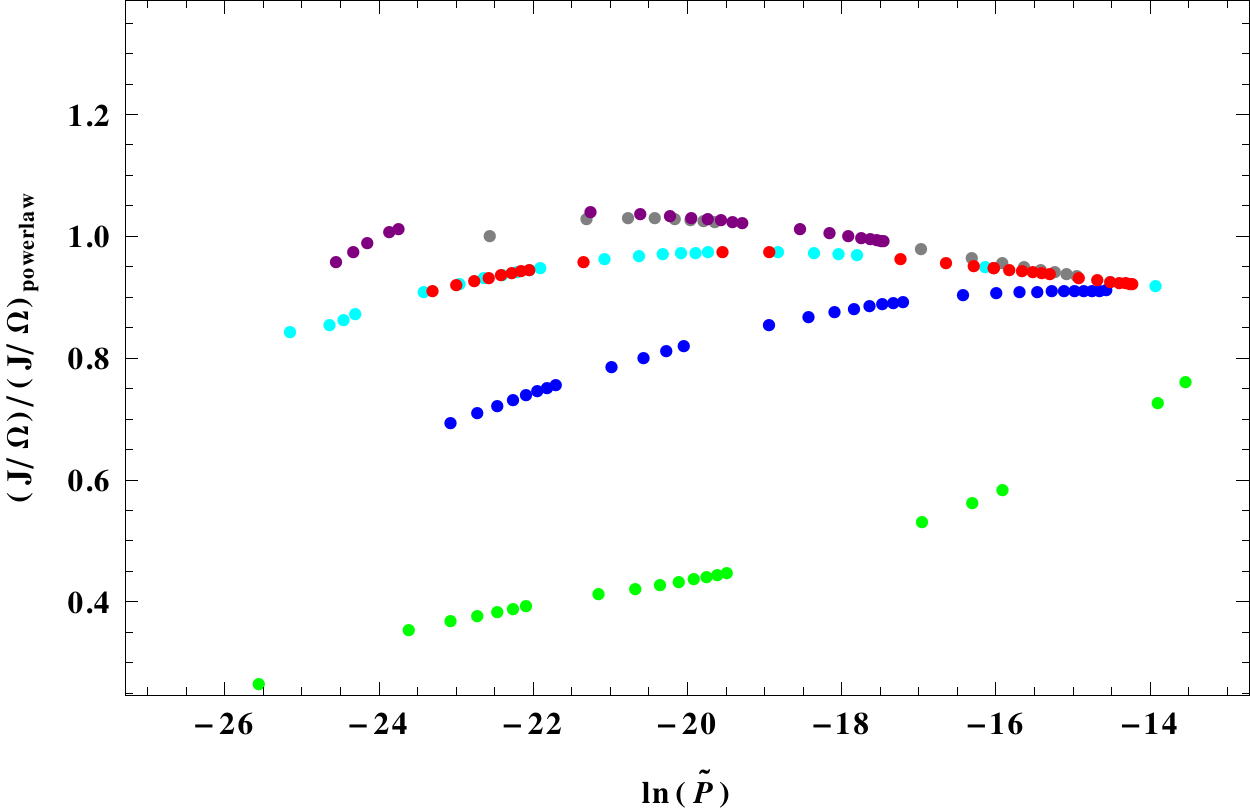}
\fi
\caption{{Values of} $J/\Omega$ with the expected leading-order power law
  taken out.}
\label{figure:Jresidual}
\end{figure}

\begin{figure}
\ifprintfigures
\includegraphics[width=3.3in]{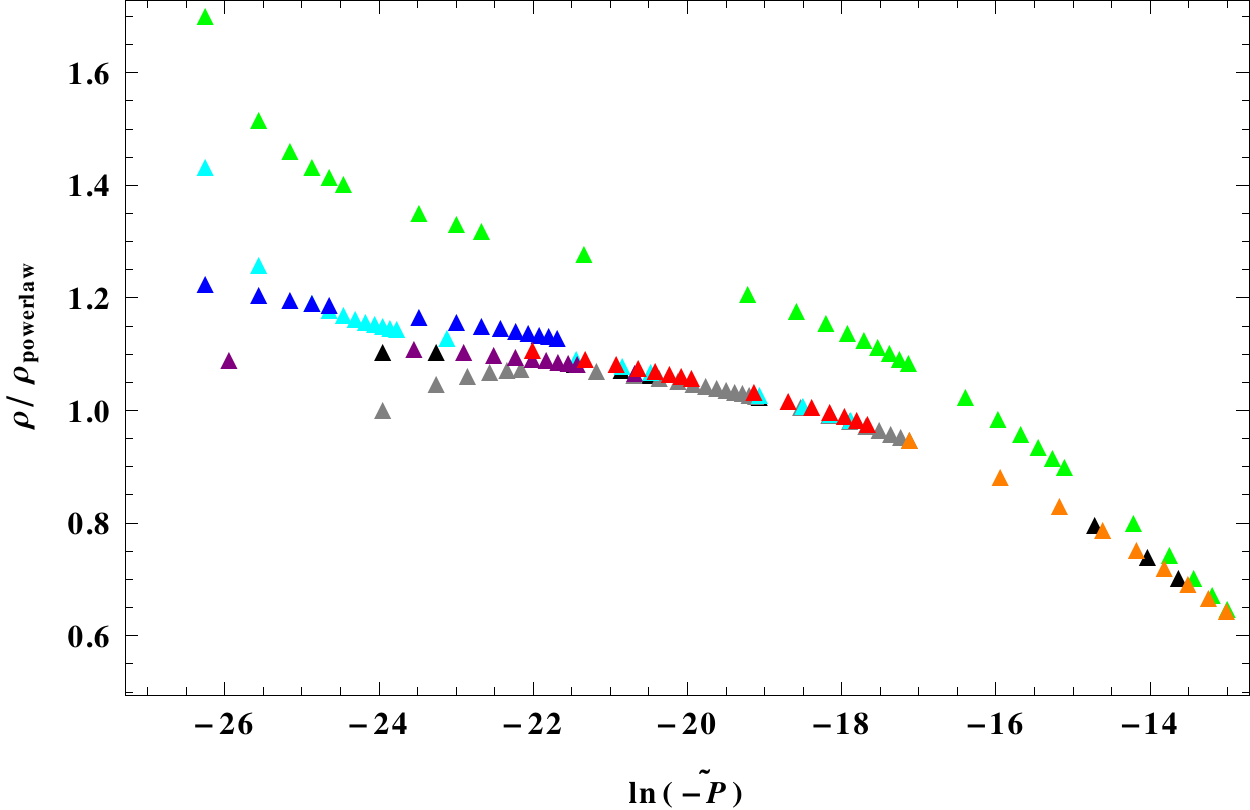}
\fi
\caption{{Maximum densities} $\rho_{\rm max}$ with the expected leading-order power
  law taken out, plotted against $\ln (-\tilde P)$.}
\label{figure:rhoresidual}
\end{figure}

\begin{figure}
\ifprintfigures
\includegraphics[width=3.3in]{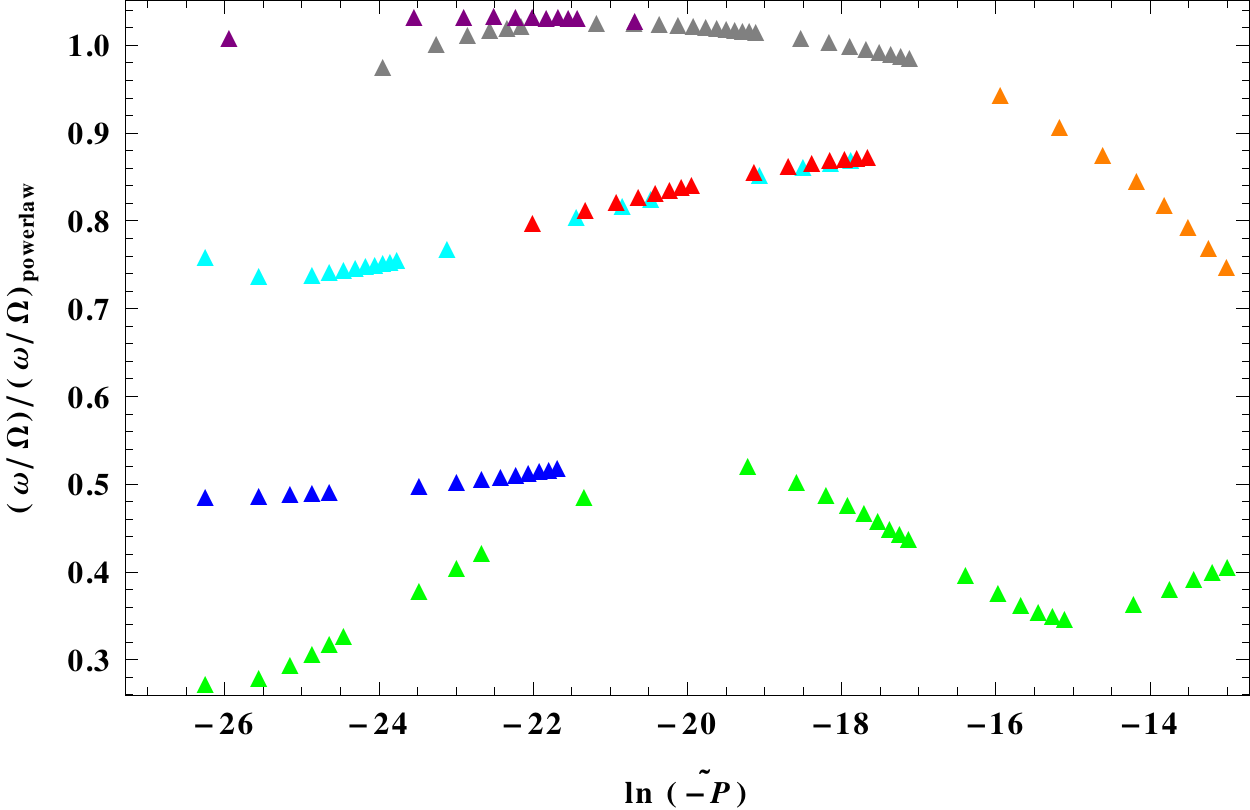}
\fi
\caption{{Values of} $\omega_{\rm max}/\Omega$ with the expected
  leading-order power law taken out.}
\label{figure:omegaresidual}
\end{figure}

To see the deviations from the leading-order power laws more clearly,
we plot the ratio of the observed quantities over the predicted
leading-order power law. For example, plotting $M/\tilde P^{\gamma_M}$
should give $\tilde F_M$. This is illustrated in
Figs.~\ref{figure:Mresidual} through \ref{figure:omegaresidual}. In
these figures we again show $\ln |\tilde P| $, that is logarithmic
distance from the black hole threshold, on the horizontal axis. In
contrast to the previous plots, where we showed all our data points,
we now restrict to $\ln|\tilde P|<-12$ to discard the data furthest from
the black hole threshold.

Fig.~\ref{figure:Mresidual} shows that the deviations of $M$ from the
leading-order power law are at the {level of a few percent} (after
removing data which are far from the black threshold), {so that we
  would not be able to distinguish these deviations from numerical
  error}.  The smallness of the deviation is consistent with $\tilde
P(\eta,\Omega)$ being well approximated by $C_0(\eta-\eta_*)$ and
$KC_1^2(\Omega_*^2-\Omega^2)$ on vertical and horizontal sequences,
respectively, and at the same time $\tilde F_M(\tilde \delta)$ being
well approximated by its leading-order term, {namely} $1$.  It is of
course possible that both approximations are violated with the
violations largely cancelling out, but this seems implausible.

For $\rho_{\rm max}$, the deviations are significantly larger
(Fig.~\ref{figure:rhoresidual}), and larger again for $J$
(Fig.~\ref{figure:Jresidual}) and $\omega_{\rm max}$
(Fig.~\ref{figure:omegaresidual}).   {It is possible that these deviations
are caused by higher-order terms in the scaling functions.}
However, for $\rho_{\rm max}$ there seems to be no clear
systematic effect, and it may be that the deviations from {unity} can be
explained by numerical error at large fine-tuning (and small scales),
and by a breakdown of our theory at $\Omega=0.3$ far from the black
hole threshold.

For $J/\Omega$ and $\omega_{\rm max}/\Omega$ there does seem to be a
systematic effect. We find that these quantities are systematically
smaller than our leading-order power law prediction with increasing
$\Omega$.

\begin{figure}
\ifprintfigures
\includegraphics[width=3.3in]{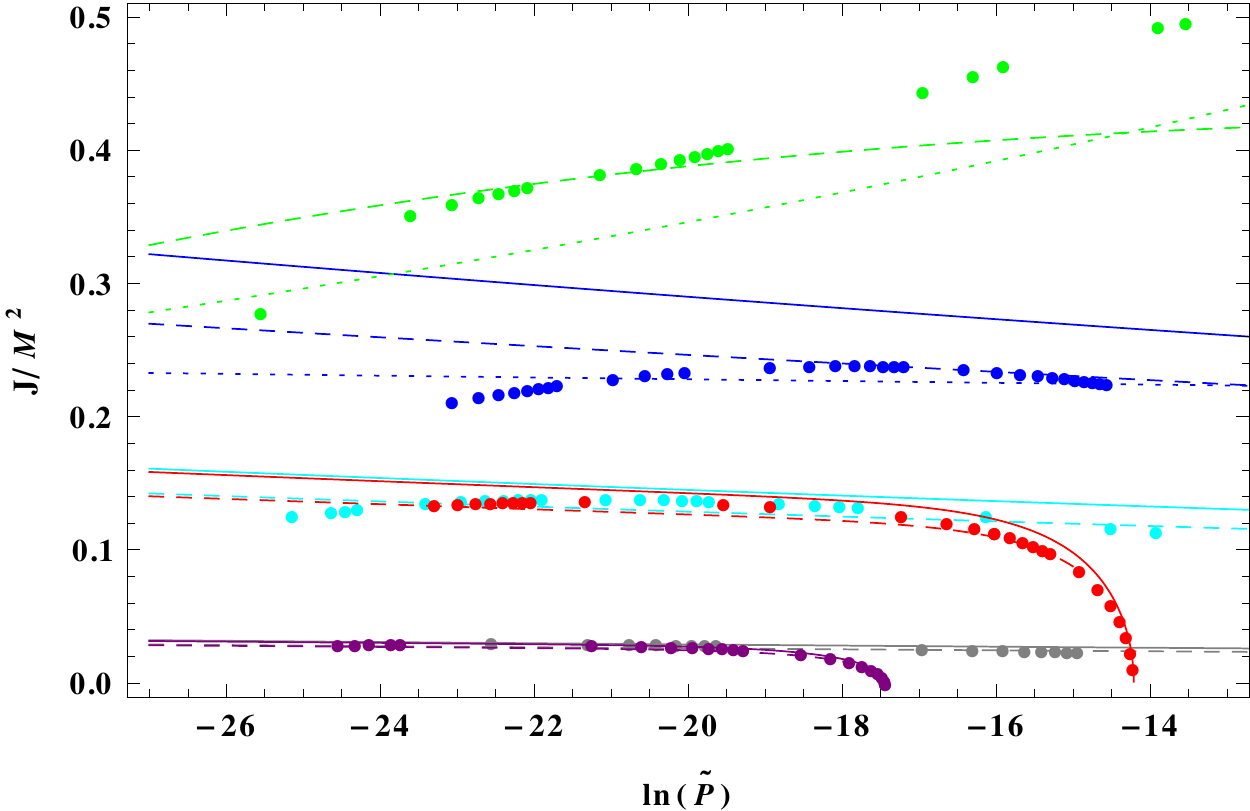}
\fi
\caption{{Numerical values of} $J/M^2$ (dots), $\tilde\delta$ as
  computed from (\ref{deltatilde}) with the theoretical value of
  $\lambda_1$ (solid lines), $\tilde\delta$ with the heuristic values
  of $\lambda_1$ for $\Omega=0.1,0.3$ (dotted lines), and the
  heuristic fit (\ref{F_fit}) for $\tilde F_{J/M^2}(\tilde\delta)$
  (dashed lines), all plotted against $\ln \tilde P$. }
\label{figure:JoM2}
\end{figure}

\begin{figure}
\ifprintfigures
\includegraphics[width=3.3in]{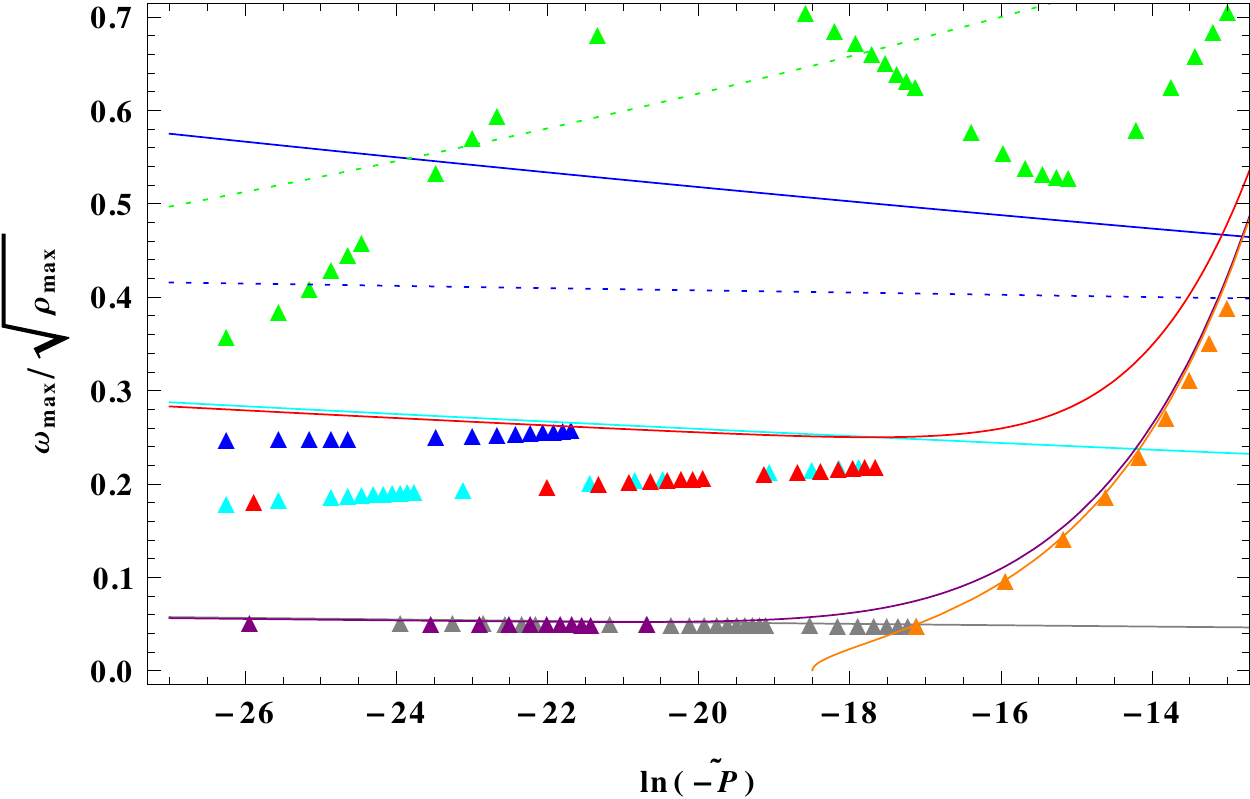}
\fi
\caption{Numerical values of $\omega_{\rm max}/\sqrt{\rho_{\rm max}}$
  (triangles), $(c_\omega/\sqrt{c_\omega})\, \tilde\delta$ with the
  theoretical value of $\lambda_1$ (solid lines) and the heuristic
  values of $\lambda_1$ for $\Omega=0.1,0.3$ (dotted lines), all
  plotted against $\ln (-\tilde P)$. For clarity, we have restricted
  the plotting range so that the solid green line (at $\delta\sim
  1.5$) is outside the frame).}
\label{figure:omegaosqrtrho}
\end{figure}

To investigate these possible systematic effects in the black hole
angular momentum (for supercritical data) and maximal central angular
velocity (for subcritical data), we now focus on the related
dimensionless quantities $J/M^2$ and $\omega_{\rm max}/\sqrt{\rho_{\rm
    max}}$. These quantities can be measured directly, unlike the
ratios between measurement and leading-order prediction shown in
Figs.~\ref{figure:Mresidual} through \ref{figure:omegaresidual}.
Moreover, $J/M^2$ and $\omega_{\rm max}/\sqrt{\rho_{\rm max}}$ should
depend on $\tilde\delta$ only. The results are shown in
Figs.~\ref{figure:JoM2} and \ref{figure:omegaosqrtrho} (colored dots
{and triangles}). The horizontal axis is again logarithmic distance to
the black hole threshold. Note that to leading order we expect
$J/M^2\simeq \tilde\delta$, and $\omega_{\rm max}/\sqrt{\rho_{\rm
    max}}\simeq (c_\omega/\sqrt{c_\rho}) \, \tilde\delta$.

\begin{figure}
\ifprintfigures
\includegraphics[width=3.3in]{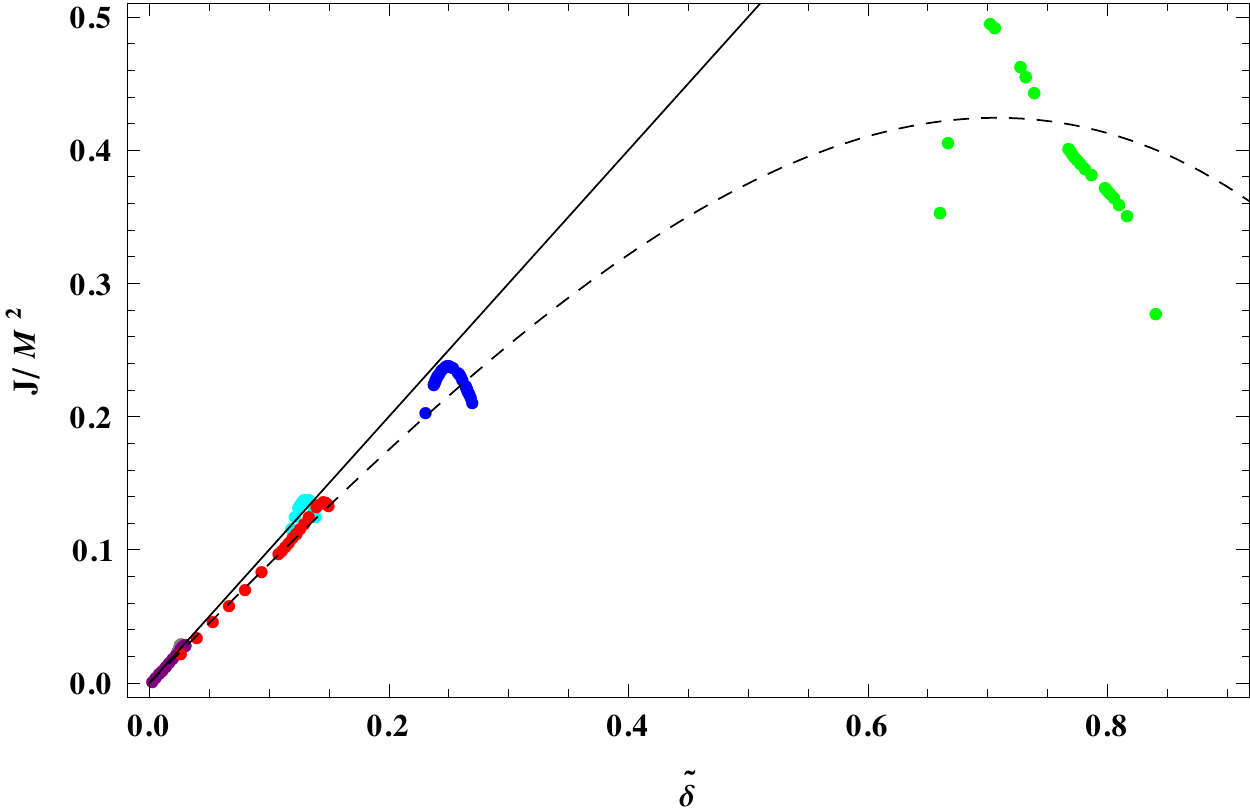}
\fi
\caption{Numerical values of $J/M^2$ (dots), $\tilde\delta$ (solid
  black line) and the heuristic fit {(\ref{F_fit})} for $\tilde
  F_{J/M^2}(\tilde\delta)$ (dashed black line), all plotted against
  $\tilde\delta$.}
\label{figure:JoM2versusdeltaplot}
\end{figure}

\begin{figure}
\ifprintfigures
\includegraphics[width=3.3in]{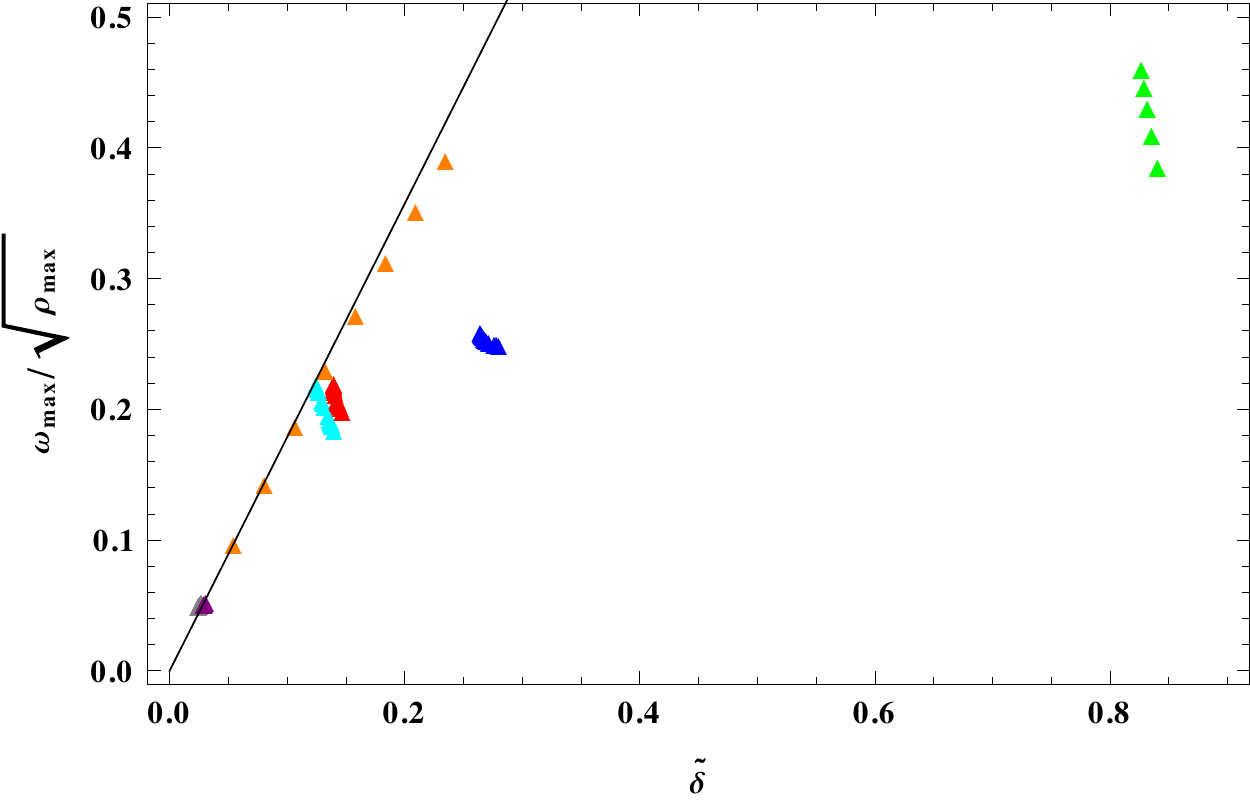}
\fi
\caption{Numerical values of $\omega_{\rm max}/\sqrt{\rho_{\rm max}}$
  (triangles) and $(c_\omega/\sqrt{c_\rho})\tilde\delta$ (solid black
  line), all plotted against $\tilde\delta$.}
\label{figure:omegaosqrtrhoversusdeltaplot}
\end{figure}

Our first observation is that both $J/M^2$ and $\omega_{\rm
  max}/\sqrt{\rho_{\rm max}}$ depend strongly on the sequence of
initial data, but very little on distance to the black hole threshold
within each sequence. Consistently with that, we find that the two
pairs of one horizontal and one vertical sequence intersecting on the
black hole threshold (grey and purple, cyan and red) are very close to
each other.

The reason for this is simply that $\epsilon {\simeq 0.015}$ is small,
as we see by also plotting $\tilde\delta$, as computed from
(\ref{deltatilde}) with the approximations (\ref{tildePvert}) through
(\ref{Qboth}) (solid colored lines). While $\tilde\delta$ increases
towards the black hole threshold (towards the left), it does so very
slowly for the degree of fine-tuning we can achieve.

As a consequence, modulo numerical error, each vertical one-parameter
sequence of initial data gives us essentially only a single point
when we plot our estimates of $\tilde F_{J/M^2}$ and $\tilde
  F_{\omega/\sqrt{\rho}}$ against $\tilde\delta$ in
  Figs.~\ref{figure:JoM2versusdeltaplot} and
  \ref{figure:omegaosqrtrhoversusdeltaplot}.  Given that $\tilde
F_{J/M^2}$ has to be an odd function in $\tilde \delta$, a reasonable
attempt at such a fit is
\begin{equation}
\label{F_fit}
\tilde F_{J/M^2}(\tilde\delta)\simeq \tilde\delta -(0.1)
\,\tilde\delta^3 -(0.6) \,\tilde\delta^5,
\end{equation}
shown as the dashed line in Fig.~\ref{figure:JoM2versusdeltaplot}. We
have compromised between using the two fitting parameters to
improve the fit of the lower-angular momentum sequences (gray, purple,
cyan, red), and a very rough fit of the two high-angular momentum
sequences (blue, green).  The resulting theoretical predictions are
also included as dashed colored lines in 
  Fig.~\ref{figure:JoM2}.

Plotting $\omega_{\rm max}/\sqrt{\rho_{\rm max}}$ against $\tilde
\delta$ does not give a consistent $\tilde
F_{\omega/\sqrt{\rho}}(\tilde\delta)$ at all, see
Fig.~\ref{figure:omegaosqrtrhoversusdeltaplot}.  The orange sequence
is well fitted by $\tilde F_{\omega/\sqrt{\rho}}=\tilde\delta$, as are
the gray/purple sequences (which effectively only contribute one
point). By contrast, the cyan/red, blue and green sequences (each
contributing effectively only one point) completely disagree with
this, even if we use the heuristic values of $\lambda_1$ for
$\Omega=0.1,0.3$.

Besides ``heuristic'' values of $\lambda_1$ and non-trivial scaling
functions, another possibility for making the theoretical values for
$J$ and $\omega_{\rm max}$ agree better with our simulations is to go
to the next order in $Q(q,p)$, that is
\begin{equation}
Q(\Omega,\eta)\simeq C_1 \,\Omega \left[1+ C_2\,(\eta-\eta_{*0})
+C_3\,\Omega^2\right].
\end{equation}
Recall that we consider $\Omega^2$ and $\eta-\eta_{*0}$ as being of
the same order of smallness. We have now expanded $Q$ to $O(\Omega^3)$
and, as before, $P$ to $O(\Omega^2)$.  In particular, in a first step
we might hope to find a value of $C_3$ that would improve the
agreement of $J/M^2$ and $\omega_{\rm max}/\sqrt{\rho_{\rm max}}$ for
$\Omega=0.1$ and $\Omega=0.3$ with the predicted values. However, the
fit for these four data points cannot be improved by a single
consistent choice of $C_1$, even roughly: the relative deviations for
$\Omega=0.3$ are not $9$ times larger than those for $\Omega=0.1$, and
the relative deviations of $J/M^2$ and $\omega_{\rm
  max}/\sqrt{\rho_{\rm max}}$ are not the same. Hence this refinement
of the theory on its own also does not appear to be promising.

Of course, all attempts to improve the agreement between numerical
results and the predicted scaling assume that the evolution still goes
through a phase II.  In accordance with our earlier discussion, it is
possible that, for $\kappa < 1/9$ and sufficiently large $\Omega$,
this is no longer the case.

\section{Conclusions}

We have studied the critical collapse of a rotating perfect fluid for
different values of the equation of state parameter $\kappa$,
generalizing our earlier results for the radiation fluid defined by
$\kappa = 1/3$. Varying $\kappa$ is interesting because, for $\kappa <
1/9$, we expect the self-similar critical solution to have two, rather
than one, unstable modes. Moreover, the second mode is controlled by
angular momentum in the initial data, and in turn controls angular
momentum in the final outcome. We have generalized the theory to
accommodate this second unstable mode in a manner that treats both
growing modes on an equal footing.  

With two growing modes, we expect qualitatively different behavior
near the black hole threshold. This is because the attracting manifold
of the critical solution, which is codimension 2 in the space of
initial data, is now only a submanifold of the black hole threshold,
which is always codimension 1. Hence by fine-tuning a generic
one-parameter family of initial data to the black-hole threshold, we
cannot set both growing modes to zero. In particular we have two key
expectations:

1. A break-down of the well-known power-law scalings at sufficient
  fine-tuning, at some minimum scale set by the position along the
  black-hole threshold (distance from the attracting manifold of the
  critical solution). 

2. The modification of the leading-order power laws by universal
  scaling functions. 

We have also performed numerical simulations of the collapse of
rotating perfect fluids.  For supercritical evolutions we measure the
black hole mass $M$ and angular momentum $J$ , while for subcritical
evolutions we introduce the maximum central rotation rate $\omega_{\rm
  max}$ as a diagnostic for critical collapse.  The latter mirrors the
scaling of the black hole angular momentum $J$ similar to how the
maximum central density $\rho_{\rm max}$ mirrors the scaling of the
black hole mass $M$ (see \cite{GarfinkleDuncan}).

We find that $M$ and $J$ as well as $\rho_{\rm max}$ and $\omega_{\rm
  max}$ are well approximated by power-laws, for $\kappa > 1/9$ as
well as $\kappa < 1/9$, at least as long as the initial rotation rate
$\Omega$ is sufficiently small.  In particular, we do not observe the
minimum scale discussed above. Unless $b_1\ll 1$ in the non-linearity
ellipse (\ref{nonlinearity}), this is not in contradiction with the
theory, basically because the second unstable mode grows much more
slowly than the first one. Observing the minimum scale would require
significantly better fine-tuning than we can currently afford.

We do observe some apparently systematic deviations from power-law
scaling for the black hole angular momentum $J$ (in supercritical
evolutions) and maximal angular rotation rate $\omega_{\rm max}$ (in
subcritical evolutions). However, our numerical observations do not
give us enough data points to make quantitative predictions about
hypothetical scaling functions, and our subcritical data do not seem
to define a consistent scaling function for $\omega_{\rm max}$.

An alternative, but at the moment purely heuristic approach, is to
assume that the Lyapunov exponent of the second unstable mode takes
heuristic values that decrease with increasing initial rotation
rate. This might be justified by nonlinear effects, but modeling these
would go far beyond our current theoretical model. We note that
\cite{ChoHLP03} found similar changes in Lyapunov exponents in the
aspherical collapse of massless scalar fields.  However, while
\cite{ChoHLP03} reported that large deformations render an aspherical
$l = 2$ mode unstable, we find that larger $\Omega$ makes the
aspherical $l = 1$ more stable. 

In summary, our numerical results are in agreement with the theory,
but so far we have been unable to make quantitative predictions about
key elements of the theory, including the scaling functions. We
believe that three different approaches could improve this in future
work.

First, the universality of the scaling functions we have fitted could
be tested by carrying out numerical evolutions of one or more other
two-parameter families of initial data. The same applies for the
universality of the conjectured heuristic values of $\lambda_1$.
{It would be interesting to} examine two-parameter families of initial data
which, at $q = 0$, have zero angular momentum but are not spherically
symmetric, and we should test the hypothesis that, for $\kappa>1/9$,
critical scaling is seen for {\em any} initial data sufficiently close
to the black-hole threshold, including rapidly spinning data.

Second, the hypothesis of a minimum scale could be tested by much
better fine-tuning to the black hole threshold. This requires higher
numerical resolution, which could be achieved either with more
sophisticated re-gridding schemes, or perhaps a gauge condition that
makes the coordinate system shrink approximately with the critical
solution, or adaptive mesh-refinement.

Third, the universal scaling functions could be determined directly by
evolving a single one-parameter family of initial data, consisting of
the critical solution with the two unstable perturbations in different
ratios, at an amplitude that is just about to become nonlinear, see
Eq.~(\ref{intermediatedata}). This will require constructing initial
data for the critical solution and its perturbations. Success in this
approach would mean that we had fewer free parameters to fit and so
would make our theory more predictive.

A key theoretical prediction is that exactly at the black hole
threshold $J/M^2$ and $\omega_{\rm max}/\sqrt{\rho_{\rm max}}$ each
take a universal limit, even if there is a minimum scale
\cite{scalingfunctions}. Evolving the initial data
(\ref{intermediatedata}) with $\alpha=\alpha_*$ would directly tell us
these universal values.  Moreover, in principle it is possible that
the universal scaling functions $F_M\to 0$ and $F_\omega\to 0$ as
$\alpha\to\alpha_*$, which means that the space of initial data that
form naked singularities could be the entire black hole threshold
(codimension one), rather than the attracting manifold of the critical
solution (codimension two).


\acknowledgements

This work was supported in part by NSF grants PHY-1402780 and 1707526
to Bowdoin College.




\end{document}